\documentclass[12pt]{amsart}

\advance\textheight\topskip
\usepackage{amsmath,amsfonts,amsthm,amssymb,amscd}
\usepackage{multicol}
\usepackage[mathscr]{eucal}
\allowdisplaybreaks
\usepackage{hyperref}
\usepackage{times}

\usepackage[curve,matrix,arrow]{xy}

\newcounter{enumcontinue}

\newcommand{\standardfootnotesize}{\footnotesize}
\newcommand{\tildeS}{\widetilde{\pureS}}
\newcommand{\catC}{\mathscr{C}}
\newcommand{\rme}{{\rm e}}
\newcommand{\rmi}{{\rm i}}
\newcommand{\dual}{^+_{}}

\newcommand{\radR}{\mathfrak{R}}
\newcommand{\End}{\mathop{\mathrm{End}}\nolimits}
\newcommand{\vectF}{F}
\newcommand{\theK}{\mathsf{K}}

\newcommand{\captionfont}[1]{\textit{\textbf{#1}}}

\newcommand{\fusion}{\circledast}
\newcommand{\falgebra}{\mathfrak{F}}
\newcommand{\algF}{\mathfrak{F}}

\newcommand{\diag}{\mathop{\mathrm{diag}}\nolimits}
\newcommand{\floor}[1]{\lfloor#1\rfloor}

\newcommand{\mfrac}[2]{\mbox{\small$\displaystyle\frac{#1}{#2}$}}
\newcommand{\ffrac}[2]{\raisebox{1pt}{\mbox{\footnotesize$\displaystyle\frac{#1}{#2}$}}}
\newcommand{\half}{\frac{1}{2}}

\newcommand{\fhalf}{\ffrac{1}{2}}

\newcommand{\bs}[1]{\boldsymbol{#1}}
\newcommand{\mat}[1]{\mathsf{#1}}  
\newcommand{\chargeC}{\mat{C}}
\newcommand{\matH}{\mat{H}}
\newcommand{\matM}{\mat{M}}
\newcommand{\eigenP}{\mat{P}}
\newcommand{\pureS}{\mat{S}}
\newcommand{\pureT}{\mat{T}}

\newcommand{\fancyS}{\boldsymbol{S}}
\newcommand{\fancyT}{\boldsymbol{T}}

\newcommand{\repPi}{\mathit{\Pi}\kern-1pt}
\newcommand{\repLambda}{\mathit{\Lambda}}
\newcommand{\repXi}{\mathit{\Xi}}
\newcommand{\repF}{\rep{F}}
\newcommand{\repK}{\rep{K}}
\newcommand{\rep}{\mathscr}  
\newcommand{\voal}{\mathcal} 
\newcommand{\algW}{\mathcal{W}}
\newcommand{\lc}{\rep{LC}}

\newcommand{\chr}[1]{\mathop{\mathrm{char}}#1}
\newcommand{\Ker}{\mathop{\mathrm{Ker}}}

\newcommand{\smatrix}[4]{\mbox{\scriptsize%
    $\displaystyle\begin{pmatrix}#1&#2\\
      #3&#4\end{pmatrix}$}}

\newcommand{\Vi}{\rep{V}_{\!-\frac{1}{8}}}
\newcommand{\Viii}{\rep{V}_{\frac{3}{8}}}

\newcommand{\dd}{\partial}
\newcommand{\SLiiZ}{SL(2,\oZ)}
\newcommand{\upperH}{\mathfrak{h}} 
\newcommand{\oC}{\mathbb{C}}

\newcommand{\oZ}{\mathbb{Z}}
\newcommand{\one}{\boldsymbol{1}}

\numberwithin{equation}{section}
\makeatletter
\@addtoreset{equation}{section}

\def\@secnumfont{\bfseries}
\def\subsubsection{\@startsection{subsubsection}{3}%
  \z@{.5\linespacing\@plus.7\linespacing}{-.5em}%
  {\normalfont\bfseries}}
\def\paragraph{\@startsection{paragraph}{4}%
  \z@\z@{-\fontdimen2\font}%
  \normalfont\bfseries}
\def\subparagraph{\@startsection{subparagraph}{5}%
  \z@\z@{-\fontdimen2\font}%
  \normalfont\bfseries}

\makeatother

\swapnumbers
\newtheorem{Thm}[subsection]{Theorem}

\newtheorem{Prop}[subsection]{Proposition}

\theoremstyle{definition}

\newtheorem{Rem}[subsection]{Remark}

\newtheorem{example}[subsubsection]{Example}

\begin{document}

\title[Nonsemisimple Verlinde Formula]{\vspace*{-4\baselineskip}
  \mbox{}\hfill\texttt{\small\lowercase{hep-th}/0306274}
  \\[2\baselineskip]
  Nonsemisimple Fusion Algebras and the\\ Verlinde Formula}

\author[Fuchs]{J.~Fuchs}%
\author[Hwang]{S.~Hwang}%

\address{\mbox{}\kern-\parindent
  jf, sh: Karlstad University, Karlstad, Sweden\hfill\mbox{}\linebreak
  \texttt{jfuchs@fuchs.tekn.kau.se}, \texttt{stephen.hwang@kau.se}}

\author[Semikhatov]{A.M.~Semikhatov}%

\address{\mbox{}\kern-\parindent  
  ams, iyt: Lebedev Physics Institute, Moscow,
  Russia\hfill\mbox{}\linebreak 
  \texttt{ams@sci.lebedev.ru}, \texttt{tipunin@td.lpi.ru}}
\author[Tipunin]{I.Yu.~Tipunin}

\begin{abstract}
  We find a nonsemisimple fusion algebra $\falgebra_p$ associated with
  each $(1,p)$ Virasoro model.  We present a nonsemisimple
  generalization of the Verlinde formula which allows us to derive
  $\falgebra_p$ from modular transformations of characters.
\end{abstract}


\maketitle
\thispagestyle{empty}




\section{Introduction}
Fusion algebras~\cite{V,kawA,CPR,jf24,Eh} describe basis-independent
aspects of operator products and thus provide essential information
about conformal field theory.  They can in principle be found by
calculating coinvariants, but the most practical derivation, which at
the same time is of fundamental importance, is from the modular
transformation properties of characters, via the Verlinde
formula~\cite{V}.  The relation between fusion and modular properties
can be considered a basic principle underlying consistency of CFT.

A fusion algebra $\algF$ is a unital commutative associative algebra
over $\oC$ with a distinguished basis (the one corresponding to the
``sectors,'' or primary fields, of the model) in which the structure
constants are nonnegative integers (we refer to this basis as the
\textit{canonical basis} of $\algF$ in what follows).

For rational CFTs, which possess semisimple fusion algebras, the
Verlinde formula is often formulated as the motto that ``the matrix
$\pureS$ diagonalizes the fusion rules.''  This involves two
statements at least.  The first is merely a lemma of linear algebra
and can be stated as follows: there exists a matrix $\eigenP$ that
relates the canonical basis in the fusion algebra to the basis of
primitive idempotents.  This property is not specific to fusion
algebras originating from conformal field theories, and in fact
applies to any association scheme~\cite{BannaiIto}; we borrow the
terminology from~\cite{BannaiIto} and call $\eigenP$ the
\textit{eigenmatrix}.  The second, nontrivial, statement contained in
the Verlinde formula is that the eigenmatrix~$\eigenP$ is related to
the matrix~$\pureS$ that represents the modular group element
$S\,{=}\,\smatrix{0}{-1}{1}{0}$ on the characters of the chiral
algebra; this relation is given by
$\eigenP\,{=}\,\pureS\,\theK_{\mathrm{diag}}$, where
$\theK_{\mathrm{diag}}$ (the denominator in the Verlinde formula) is a
diagonal matrix whose entries are the inverse of the distinguished row
of the $\pureS$ matrix.  With $\eigenP$ expressed this way, we arrive
at the statement that the matrices of the regular representation of
the fusion algebra are diagonalized by the $\pureS$ matrix.

This cannot apply to nonsemisimple fusion algebras, however, for which
the regular representation matrices cannot be diagonalized.  The
relation between modular transformations and the structure of
nonsemisimple fusion algebras is therefore more difficult to identify,
which considerably complicates attempts to build a nonsemisimple
Verlinde formula.

Nonsemisimple fusion algebras are expected to arise in logarithmic
models of conformal field
theory~\cite{Gurarie,GK1,GK2,Roh,KoMa,KoLe,F96,G-alg,FFHST}, where
irreducible representations of the chiral algebra allow nontrivial
(indecomposable) extensions.  In what follows, we generalize the
Verlinde formula and derive nonsemisimple fusion algebras for the
series of $(1,p)$ Virasoro models with integer $p\,{\geq}\,2$.

The $(1,p)$ models provide an excellent illustration of complications
involved in generalizing the Verlinde formula to the nonsemisimple
case.  Unlike the $(p',p)$ models with coprime $p',p\,{\geq}\,2$, the
$(1,p)$ model is defined not as the cohomology, but as the kernel of a
screening, and the first question that must be answered in
constructing its fusion, as well as the fusion beyond minimal models
in general,~is:
\begingroup\renewcommand{\labelenumi}{\textup{Q\theenumi:}}
\begin{enumerate}
\item\label{which-algebra}How to organize the Virasoro representations
  into a finite number of families?  That is, which chiral algebra,
  extending the Virasoro algebra, is to be used to classify
  representations?%
  \setcounter{enumcontinue}{\theenumi}
\end{enumerate}
\noindent
Assuming that such an algebra has been chosen, the fusion algebra can
in principle be derived using different means, e.g., by directly
finding coinvariants (if, against expectations, this is feasible).
Another possibility is via a Verlinde-like formula, starting with
\textit{characters} of representations of the chosen chiral algebra.
Compared to the semisimple case, the basic problems with constructing
an analogue of the Verlinde formula are then as follows.
\begin{enumerate}\setcounter{enumi}{\theenumcontinue}
  
\item\label{involve-tau}The matrices implementing modular
  transformations of the characters of chiral algebra representations
  involve the modular parameter~$\tau$ and \textit{do not therefore
    generate a finite-di\-mensional representation of~$\SLiiZ$}.  How
  to extract a $\tau$-independent matrix $\pureS$ representing
  $S\,{\in}\,\SLiiZ$ on characters?
  
\item\label{which-canonical} With fusion matrices that are not
  simultaneously diagonalizable, it is not a priori known which
  ``special'' (instead of diagonal) matrix form is to be used in a
  Verlinde-like formula.  In other words, how to define the
  eigenmatrix $\eigenP$ that performs the transformation to a
  ``special'' form in a nonsemisimple fusion algebra?
  
\item\label{what-K} Assuming the matrix $\pureS$ is known and the
  matrix $\eigenP$ that performs the transformation to the chosen
  ``special'' basis has been selected, \textit{how are $\pureS$ and
    $\eigenP$ related}?

\setcounter{enumcontinue}{\theenumi}
\end{enumerate}
\noindent
The most nontrivial part of the nonsemisimple generalization of the
Verlinde formula is the answer to~Q\ref{what-K}.  We also note that
with a chiral algebra chosen in~Q\ref{which-algebra}, we face yet
another problem of a ``nonsemisimple'' nature, originating in the
structure of the category of representations of the chosen chiral
algebra:

\begin{enumerate}\setcounter{enumi}{\theenumcontinue}
\item\label{indecomposable}With \textit{indecomposable}
  representations of the chiral algebra involved, how many generators
  should the fusion algebra have?  More specifically, whenever there
  is a nonsplittable exact sequence
  $0\,{\to}\,\rep{V}_0\,{\to}\,\rep{R}\,{\to}\,\rep{V}_1\,{\to}\,0$,
  should there be fusion algebra generators corresponding to one
  (i.e., $\rep{R}$), two (i.e., $\rep{V}_0$ and~$\rep{V}_1$), or three
  representations?  (This becomes critical, e.g., when $\rep{V}_0$
  corresponds to the unit element of the fusion algebra,
  cf.~\cite{GK2}).
\end{enumerate}
\endgroup

We also note the following complications that are already apparent in
nonunitary semisimple fusion (see the relevant remarks
in~\cite{Gannon}), but become more acute for nonsemisimple fusion:
\begingroup\renewcommand{\labelenumi}{\textup{R\theenumi:}}
\begin{enumerate}
\item\label{nonsymmetric}Whenever the $\pureS$ matrix is not
  symmetric, the sought generalization of the Verlinde formula is
  sensitive to the choice between $\pureS$ and $\pureS^{t}$. This is
  essential, in particular, in selecting the distinguished row/column
  of $\pureS$ whose elements determine eigenvalues of the fusion
  matrices (the denominator of the Verlinde formula).\footnote{In
    addition, it becomes essential whether a representation or an
    \textit{anti}representation of $\SLiiZ$ is considered as the
    modular group action (in most of the known semisimple examples,
    this point can safely be ignored).}
  
\item\label{distinguished-row} The sector with the minimal conformal
  weight is different from the vacuum sector.  It is therefore
  necessary to decide which of these two distinguished sectors is to
  play the ``reference'' role in the Verlinde formula.  (That is, as a
  continuation of the previous question, the distinguished row of the
  $\pureS$ matrix to be used in the denominator of the Verlinde
  formula must be identified properly).\enlargethispage{\baselineskip}

\end{enumerate}
\endgroup

In answering Q\ref{indecomposable}, one must be aware that fusion
algebras only provide a ``coarse-grained'' description of conformal
field theory, and there can be several degrees of neglecting the
details.  The concept of nonsemisimple fusion advocated
in~\cite{GK1,GK2} aims at accounting for the ``fine'' structure given
by the different ways in which simple (irreducible) modules can be
arranged into indecomposable representations. (Such a detailed fusion
will be needed, e.g., in studying boundary conditions in conformal
field theory models and for a proper interpretation of modular
invariants.)  In that setting, a natural basis in the fusion algebra
would be given by \textit{all} indecomposable representations (the
irreducible ones included).\footnote{The $p\,{=}\,2$ fusion
  in~\cite{GK1,GK2} is ``intermediate'' in that not all of the
  indecomposable representations are taken into account.  But it is
  certainly sufficient for extracting the coarser, ``$K_0$''-fusion
  that follows from Theorem~\ref{Thm:fusion} below for $p=2$.}
A~coarser description is to think of the fusion algebra as the
Grothendieck ring of the representation category of the chiral
algebra, i.e., as the $K_0$ functor, not distinguishing between
different compositions of the same subquotients.  \textit{This fusion
  is sufficient for the construction of the generalized Verlinde
  formula}.  Indeed, the appropriately generalized Verlinde formula
should relate the matrix $\pureS$ that represents $S\,{\in}\,\SLiiZ$
on a collection of characters of the chiral algebra to the fusion
algebra structure constants.  But the character of an indecomposable
representation $\rep{R}$ is \textit{the sum of the characters} of its
simple subquotients, independently of how the algebra action ``glues''
them into~$\rep{R}$.  Therefore, for the fusion functor defined for
the purpose of constructing the generalized Verlinde formula, an
indecomposable representation $\rep{R}$ as in Q\ref{indecomposable} is
indistinguishable from the direct sum of $\rep{V}_0$ and $\rep{V}_1$
(as well as from $\rep{R}'$ in
$0\,{\to}\,\rep{V}_1\,{\to}\,\rep{R}'\,{\to}\,\rep{V}_0\,{\to}\,0$).
In other words, the element of the fusion algebra corresponding
to~$\rep{R}$ is the sum of those corresponding to~$\rep{V}_0$
and~$\rep{V}_1$.  In this paper, we only deal with this particular
concept of fusion that corresponds to the $K_0$ functor.

Thus, the number of elements in a basis of the fusion algebra
associated with a collection $\{\rep{V}_j,\rep{R}_i\}$ of chiral
algebra representations must be given by the number of all
\textit{simple subquotients} of all the indecomposable
representations~$\rep{R}_i$ and all simple $\rep{V}_j$ (with each
irreducible representation occurring just once).  But the fact that no
linearly independent elements of the fusion algebra correspond to
indecomposable representations does \textit{not} mean that
``nonsemisimple effects'' are neglected: the existence of a nontrivial
extension of any two representations $\rep{V}_0$ and $\rep{V}_1$
already makes the fusion algebra nonsemisimple, giving rise to all of
the problems listed~above.

The answer to Q\ref{which-algebra} can be extracted from the
literature~\cite{K-first,GK2}: we take the maximal local subalgebra in
the (nonlocal) chiral algebra that is naturally present in the $(1,p)$
model. This W algebra, denoted by~$\algW(p)$ for brevity, has $2p$
irreducible representations in the $(1,p)$ model.

As regards Q\ref{involve-tau}, the answer amounts to the use of
\textit{matrix} automorphy ``factors,'' as explained below
(cf.~\cite{EhSk}).  The answer to~Q\ref{which-canonical} is related to
the structure of associative algebras~\cite{Pierce} and, once a
canonical basis is fixed, to nonsemisimple generalizations of some
notions from the theory of association
schemes~\cite{BannaiIto}. Any finitely generated
associative algebra $\algF$ (with a unit) is the vector-space sum of a
distinguished ideal~$\radR$, called the \textit{radical} (the
intersection of all maximal left ideals, or equivalently, of all
maximal right ideals), and a semisimple algebra (necessarily
isomorphic to a direct sum of matrix algebras over division algebras
over the base field)~\cite{Pierce}.  This implies that in any
commutative associative algebra, there is a basis
\begin{equation*}
  (e_A, w_\alpha),\quad A=1,\dots,n',\quad\alpha=1,\dots,n''
\end{equation*}
(with $n'{+}n''\,{=}\,n\,{=}\,\dim \algF$), composed of primitive
idempotents $e_A$ and elements $w_\alpha$ in the radical.  In the
semisimple case, the radical is zero, and ``diagonalization of the
fusion'' can equivalently be stated as the transformation to the basis
$(\lambda_1 e_1,\dots,\lambda_n e_n)$ of ``rescaled idempotents,''
where $\lambda_a$ are scalars read off from the distinguished row of
the~$\pureS$~matrix (the row corresponding to the vacuum
representation).  Let \ $X_I$, $I\,{=}\,1,\dots,n$, denote the
elements of the canonical basis in the fusion algebra.  Even for
semisimple algebras, it is useful to distinguish between
the~$\pureS$~matrix that transforms the canonical basis $X_\bullet$ to
the basis $(\lambda_1 e_1,\dots,\lambda_n e_n)$ and the matrix
$\eigenP$ that transforms the canonical basis to the basis of
primitive idempotents, even though $\pureS$ and $\eigenP$ are related
by multiplication with a diagonal matrix.  In the nonsemisimple case,
the \textit{eigenmatrix} $\eigenP$ that maps the canonical basis to
the basis consisting of primitive idempotents and elements in the radical,
\begin{equation*}
  \begin{pmatrix}
    X_1\\[-4pt]
    \vdots\\[-2pt]
    X_n
  \end{pmatrix}
  {}={}\eigenP
  \begin{pmatrix}
    e_A\\[5pt]
    w_\alpha
  \end{pmatrix}\!,
\end{equation*}
is related to the $\pureS$ matrix in a more complicated way.  The
resolution of~Q\ref{what-K}, which is the heart of the nonsemisimple
Verlinde formula, is the construction, from the entries of~$\pureS$,
of a (nondiagonal) \textit{interpolating matrix}~$\theK$ (which plays
the role of the denominator in the Verlinde formula) such that
\begin{equation*}
  \eigenP=\pureS\,\theK.
\end{equation*}

The points raised in~R\ref{distinguished-row} and~R\ref{nonsymmetric}
can be clarified as follows.  The rows and columns of $\pureS$ are
labeled by chiral algebra representations in the model under
consideration. The $\pureS$ matrix has a distinguished row that
corresponds to the vacuum representation and a distinguished column
that corresponds to the minimum-dimension representation of the chiral
algebra (the entries in this column are related to the asymptotic form
of the characters labeled by the respective rows of~$\pureS$).  The
columns of the $\eigenP$ matrix are labeled by the elements
$(e_A,w_\alpha)$ of the basis consisting of primitive idempotents and
elements in the radical, and its rows correspond to elements of the
canonical basis in the fusion algebra; the distinguished row
of~$\eigenP$ then corresponds to the unit element of the algebra. (The
choice of rows vs.\ columns in $\eigenP$, $\pureS$, and other matrices
is of course conventional, but the replacement
rows$\,\,{\leftrightarrow}\,\,$columns must be made consistently with 
other transpositions and change of the order in matrix multiplication.)

We now summarize our strategy to construct the $(1,p)$ fusion via a
nonsemisimple generalization of the Verlinde formula and also describe
the contents of the paper:
\begin{enumerate}
\item In the $(1,p)$ model, we identify the maximal local algebra
  $\algW(p)$ as the chiral algebra of the model.  There then exist
  only $2p$ irreducible $\algW(p)$ representations in the model,
  which solves~Q\ref{which-algebra}. (The algebra is introduced in
  Sec.~\ref{W(p)}, and its category of representations is described in
  Secs.~\ref{W-reps} and~\ref{rep-exts}.)
  
\item We then evaluate the $2p$ characters $\bs{\chi}$ of these
  representations and find ($\tau$-dependent) $2p\,{\times}\,2p$
  matrices $\bs{J}(\gamma,\tau)$ such that
  $\bs{\chi}(\gamma\tau)\,{=}\,\bs{J}(\gamma,\tau)\,\bs{\chi}(\tau)$
  for $\gamma\!\in\!\SLiiZ$. (The characters are evaluated in
  Sec.~\ref{chars} and their modular transformation properties are
  derived in Sec.~\ref{sec:fancy-S}.)
  
\item Next, we find a $2p\,{\times}\,2p$ automorphy ``factor''
  $j(\gamma,\tau)$, satisfying the cocycle condition, such that
  $\gamma\,{\mapsto}\,\rho(\gamma)
  \,{=}\,j(\gamma,\tau)\bs{J}(\gamma,\tau)$ is a
  \textit{representation} of~$\SLiiZ$. This solves~Q\ref{involve-tau}
  (Secs.~\ref{automorphy} and~\ref{representation}) and gives the
  $\pureS$ matrix (Sec.~\ref{representation1p}).
  
\item From the entries of the distinguished row
  of~$\pureS\,{=}\,\rho(S)$, we build the interpolating matrix~$\theK$
  and use it to construct the eigenmatrix~$\eigenP$ of the fusion
  algebra as $\eigenP\,{=}\,\pureS\,\theK$. This solves~Q\ref{what-K}
  (Sec.~\ref{sec:StoP}).
  
\item From the eigenmatrix $\eigenP$, we uniquely reconstruct the
  fusion algebra~$\falgebra_p$ in the canonical basis whose elements
  are labeled by the rows of~$\eigenP$, via a recipe that involves
  answering~Q\ref{which-canonical} (Sec.~\ref{sec:find-fusion}).

\end{enumerate}

\textit{For the impatient}, we here present the answer for the structure 
constants expressed through the entries of the $\pureS$ matrix: arranged 
into matrices $N_I$ in the standard way, the structure constants of the 
fusion algebra are given by $N_I\,{=}\,\pureS \mathscr{O}_I\pureS^{-1}$, 
where $\pureS\,{=}\,\rho(S)$ acts in a finite-dimensional (in $(1,p)$
models, $2p$-dimensional) representation of $\SLiiZ$ and
$\mathscr{O}_I\,{=}\,\mathscr{O}_{I0}\,{\oplus}\,\mathscr{O}_{I1}\,
{\oplus}\dots{\oplus}\,\mathscr{O}_{I,p-1}$ are block-diagonal matrices 
with the $2\,{\times}\,2$ blocks given by $\smash[b]{\mathscr{O}_{I0}
  \,{=}\,\diag\bigl(\frac{S_{I}^{\ 1}}{S_{\Omega}^{\ 1}}, \frac{S_{I}^{\ 
      2}}{S_{\Omega}^{\ 2}}\bigr)}$ and
\begin{multline*}
  \mathscr{O}_{Ij}=
    \ffrac{1}{S_{\Omega}^{\ 2j+1}-S_{\Omega}^{\ 2j+2}}
  \times{}\\*[4pt]
  \mbox{\addtolength{\arraycolsep}{-12pt}
    \standardfootnotesize$\displaystyle\begin{pmatrix}
      S_{\Omega}^{\ 2j+1}S_{I}^{\ 2j+1}
      \,{-}\,2S_{\Omega}^{\ 2j+2}S_{I}^{\ 2j+1}
      \,{+}\,S_{\Omega}^{\ 2j+2}S_{I}^{\ 2j+2}&
      -S_{\Omega}^{\ 2j+2}S_{I}^{\ 2j+1}
      \,{+}\,S_{\Omega}^{\ 2j+1}S_{I}^{\ 2j+2}\\[8pt]
      S_{\Omega}^{\ 2j+2}S_{I}^{\ 2j+1}
      \,{-}\,S_{\Omega}^{\ 2j+1}S_{I}^{\ 2j+2}&
      S_{\Omega}^{\ 2j+1}S_{I}^{\ 2j+1}
      \,{-}\,2S_{\Omega}^{\ 2j+1}S_{I}^{\ 2j+2}
      \,{+}\,S_{\Omega}^{\ 2j+2}S_{I}^{\ 2j+2}
    \end{pmatrix}\!,$}
\end{multline*}
where $j\,{=}\,1,\dots,p{-}1$ and $S_{I}^{\ J}$ with
$I,J\,{=}\,1,2,\dots,2p$ are entries of the $\pureS$ matrix, with
$\pureS_{\Omega}^{\ \bullet}$ being its row corresponding to the
vacuum representation.  Thus written, these formulas may seem messy
(and the labeling of $S_I^{\ J}$ involves a convention on ordering the
representations in accordance with their linkage classes), but they in
fact have a clear structure (Eqs.~\eqref{build-P},
\eqref{vac-row}\,--\,\eqref{build-K},
and~\eqref{M_I}\,--\,\eqref{N-from-M}), to be explained in what
follows.  The resulting $(1,p)$ fusion algebra is given in
Theorem~\ref{Thm:fusion}.  A posteriori, it turns out to have positive
integral coefficients, although we do not derive the proposed recipe
for the generalized Verlinde formula from first principles such that
this property would be guaranteed in advance.

\section{The maximal local W algebra in the $(1,p)$ model}
\subsection{Energy-momentum tensor, screening operators, and
  resolutions}\label{sec:aa} 
For the $(p',p)$ minimal Virasoro models with coprime $p',p\,{\geq}\,2$, 
the Kac table of size $(p'{-}1)$\linebreak[0]$%
{\times}\,(p{-}1)$ (after suitable identifications of boxes) contains
those modules that do not admit nontrivial extensions among
themselves.  The extended Kac table of size $p'\,{\times}\, p$ then
corresponds to a logarithmic extension.  The Kac table is selected as
the cohomology, and the extended Kac table as the kernel, of an
appropriate screening.  We consider the models with $p'\,{=}\,1$,
where the Kac table is empty, while the extended Kac table has size
$1\,{\times}\,p$, with its boxes corresponding to Virasoro
representations $\rep{V}_s$, $s\,{=}\,1,\dots,p$.  Similarly to the
logarithmically extended $(p',p)$ models, the $(1,p)$ model is also
defined as the kernel of the corresponding screening operator (this
does not automatically yield its chiral algebra, however, which has
then to be found, see below).

The conformal dimensions (weights) of the primary fields corresponding
to the irreducible modules $\rep{V}_s$, $s\,{=}\,1,\dots,p$, are given
by $\Delta(1,s)$, where for future use we define
\begin{equation*}
  \Delta(r,s):=\ffrac{r^2-1}{4}\,p + \ffrac{s^2-1}{4p} + \ffrac{1-rs}{2}.
\end{equation*}
In the free-field realization through a scalar field $\varphi(z)$ with
the OPE
\begin{equation*}
  \varphi(z)\,\varphi(w)= \log(z{-}w),
\end{equation*}
the corresponding primary fields are represented by the vertex
operators $\,\rme^{j(1,s)\varphi}$, where
\begin{equation*}
  j(r,s):=\ffrac{1-r}{2}\,\alpha_+ + \ffrac{1-s}{2}\,\alpha_-
\end{equation*}
with
\begin{equation*}
  \alpha_+=\sqrt{2p},\qquad\alpha_-=-\sqrt{\ffrac{2}{p}}\,.
\end{equation*}
Because $p\alpha_- \,{=}\, {-}\alpha_+$, we have
$j(r,s{+}np)\,{=}\,j(r{-}n,s)$, $n\,{\in}\,\oZ$.  The
energy-mom\-en\-tum tensor is given by
\begin{equation}\label{eq:the-Virasoro}
  T=\fhalf\,\dd\varphi\,\dd\varphi+\ffrac{\alpha_0}{2}\,\dd^2\varphi
\end{equation}
(here and in similar formulas below, normal ordering is implied in the
products), where $\alpha_0\,{=}\,\alpha_+ \,{+}\, \alpha_-$, and the
central charge is $c\,{=}\,13\,{-}\,6(p\,{+}\,\frac1p)$.  There then
exist two screening operators
\begin{equation*}
  S_+=\oint \rme^{\alpha_+\varphi},\qquad
  S_-=\oint \rme^{\alpha_-\varphi},
\end{equation*}
satisfying $[S_{\pm},T(z)]\,{=}\,0$.

Let $\rep{F}_{j(r,s)}$ denote the Fock module generated from (the
state corresponding to) the vertex operator $\rme^{j(r,s)\varphi}$ by
elements of the Heisenberg algebra generated by the modes of the
current~$\dd\varphi$.  Set~$\rep{F}_{[s]}\,{=}\,\rep{F}_{j(1,s)}$, and
let the corresponding Feigin--Fuchs module over the Virasoro
algebra~\eqref{eq:the-Virasoro} be denoted by the same symbol.  For
each~$1\,{\leq}\, s\,{\leq}\, p{-}1$, \ $\rep{F}_{[s]}$ is included
into the acyclic Felder complex
\begin{equation}\label{Felder}
  \dots\,\leftarrow\,\rep{F}_{[s-2p]}\,\xleftarrow{S_-^{p-s}}\,
  \rep{F}_{[-s]}\,\xleftarrow{S_-^s}
  \rep{F}_{[s]}\,
  \,\xleftarrow{S_-^{p-s}}\rep{F}_{[-s+2p]}\,
  \,\xleftarrow{S_-^s}\rep{F}_{[s+2p]}\,\leftarrow\,\dots,
\end{equation}
where $\rep{F}_{[\pm s+2np]}\,{=}\,\rep{F}_{j(1-2n,\pm s)}$.

We define a (nonlocal) algebra $\voal{A}(p)$ as the kernel
\begin{equation*}
  \voal{A}(p):=\Ker S_-\!\bigm|_{\repF}
\end{equation*} 
of the $S_-$ screening on the direct sum 
\begin{equation*} 
  \repF:=\bigoplus_{\substack{r\in\oZ\\ s=1,\dots,p}}\rep{F}_{j(r,s)}
\end{equation*}
of Fock modules. The algebra $\voal{A}(p)$ is generated by
\begin{equation*}
  a^-:=\rme^{-\frac{\alpha_+}{2}\,\varphi}\qquad\text{and}\qquad
  a^+:=[S_+, a^-]
\end{equation*} 
and is therefore determined by the lattice $\half\alpha_+\oZ$. It is
slightly nonlocal: the scalar products of lattice vectors are in
$\half\oZ$.

\subsection{The maximal local algebra}\label{W(p)}
We next consider the W algebra that is the \textit{maximal local
  subalgebra} in~$\voal{A}(p)$ and use the notation $\algW(p)$ for
it for brevity.  It is generated by the three currents $W^-$, $W^0$,
and $W^+$ given by
\begin{equation*}
  W^-(z):=\rme^{-\alpha_+\varphi}(z),\quad\;
  W^0(z):=[S_+,W^-(z)],\quad\;
  W^+(z):=[S_+,W^0(z)].
\end{equation*}
We note that $W^0$ is a (free-field) descendant of the identity
operator, while $W^+$ is a descendant of $\rme^{\alpha_+\varphi}$.
The fields $W^-$, $W^0$, and $W^+$ are Virasoro primaries; their
conformal dimensions are given by $2p{-}1$.


\begin{example}  
  For $p\,{=}\,3$, the $\algW(3)$ generators are given by
  $W^-\,{=}\,\rme^{-\sqrt{6}\varphi}$,
  \begin{multline*}
    W^0=\fhalf\,\dd^3\varphi\,\dd^2\varphi +
    \ffrac{1}{4}\,\dd^4\varphi\,\dd\varphi +
    \ffrac{3}{2}\sqrt{\ffrac{3}{2}}
    \,\dd^2\varphi\,\dd^2\varphi\,\dd\varphi
    + \sqrt{\ffrac{3}{2}}\,\dd^3\varphi\,\dd\varphi\,\dd\varphi\\*
    {}+ 3\,\dd^2\varphi\,\dd\varphi\,\dd\varphi\,\dd\varphi +
    \ffrac{3}{5}\,\sqrt{\ffrac{3}{2}}
    \,\dd\varphi\,\dd\varphi\,\dd\varphi\,\dd\varphi\,\dd\varphi +
    \ffrac{1}{20\sqrt{6}}\,\dd^5\varphi,
  \end{multline*}
  and
  \begin{multline*}
    W^+=\Bigl( -\sqrt{\ffrac{3}{2}}\,\dd^4\varphi - 39\,\dd^2\varphi
    \,\dd^2\varphi
    + 18\,\dd^3\varphi\,\dd\varphi\\*
    + 12\sqrt{6}\,\dd^2\varphi \,\dd\varphi \,\dd\varphi
    - 18\,\dd\varphi \,\dd\varphi \,\dd\varphi \,\dd\varphi\Bigr)
    \rme^{\sqrt{6}\varphi}
  \end{multline*}
  (in the last formula, despite the brackets introduced for
  compactness of notation, the nested normal ordering is from right to
  left, e.g.,
  $\dd^2\varphi(\dd\varphi(\dd\varphi(\rme^{\sqrt{6}\varphi})))$).
\end{example}

\subsection{$\algW(p)$ representations}\label{W-reps} 
\begin{figure}[tb]
  \mbox{}~\quad~
  \xymatrix@=6pt{%
    &&&&&&&&\\
    &&&&&&&{\bullet}
    \ar@{}[-1,0]|(.8){\displaystyle\rep{F}_{j(1,s)}}
    \ar[3,-2]_{a^-_{\frac{2s-3p}{4}}}
    \ar[3,2]^{a^+_{\frac{2s-3p}{4}}}\\
    &&&&&&&&\\
    &&&&&&&&\\
    &&&&&{\circ}\ar[5,-2]_{a^-_{\frac{2s-5p}{4}}}
    \ar@{}[-1,-2]|(.6){\displaystyle\rep{F}_{j(2,s)}}
    &&&&{\circ}\ar[5,2]^{a^+_{\frac{2s-5p}{4}}}\\
    &&&&&&&&&&&&\\
    &&&&&&&&&&&&&\\
    &&&&&&&&&&&&\\
    &&&&&&&&&&&&&\\
    &&&{\bullet}\ar[7,-2]_{a^-_{\frac{2s-7p}{4}}}
    \ar@{}[-1,-2]|(.7){\displaystyle\rep{F}_{j(3,s)}}
    &&&&&&&&{\bullet}\ar[7,2]^{a^+_{\frac{2s-7p}{4}}}\\
    &&&&&&&&&&&&&&&&\\
    &&&&&&&&&&&&&&&&\\
    &&&&&&&&&&&&&&&&\\
    &&&&&&&&&&&&&&&&\\
    &&&&&&&&&&&&&&&&\\
    &&&&&&&&&&&&&&&&\\
    &{\circ}\ar@{.}[];[]+<-3pt,-18pt>
    \ar@{}[0,-1]-<20pt,0pt>|(.6){\displaystyle\rep{F}_{j(4,s)}}
    &&&&&&&&&&&&{\circ}
    \ar@{.}[];[]+<3pt,-18pt>
    }
  \caption[The $\repLambda$ and $\repPi$ modules]{\small\captionfont{The
      $\repLambda$ and $\repPi$ modules.}  Filled (open) dots denote
    Virasoro representations that are combined in~$\repLambda(s)$
    (respectively, $\repPi(s)$).  The $\voal{A}(p)$ generators $a^+$
    and $a^-$ act between these representations with noninteger modes,
    but $\algW(p)$ generators (not indicated) connecting dots of
    the same type are integer-moded.}
   \label{fig:LambdaPi}
\end{figure}
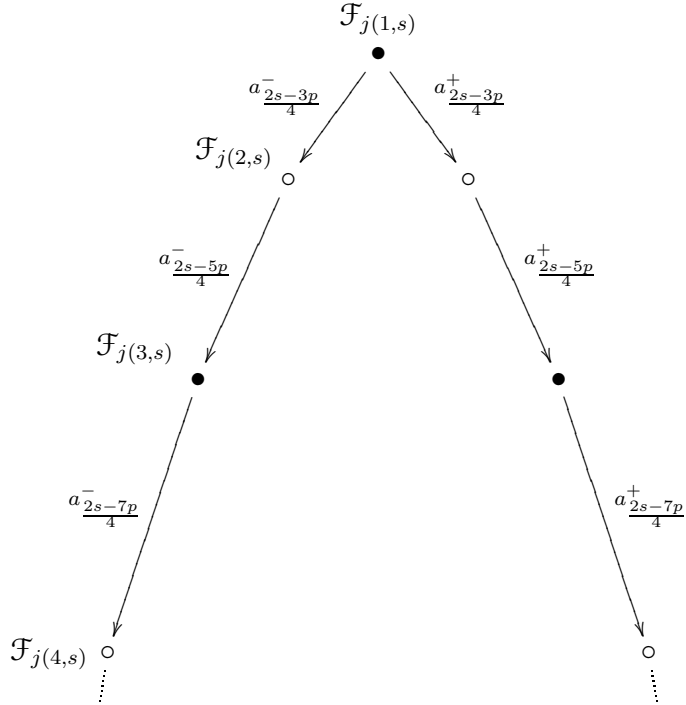

The $\algW(p)$ generators change the ``momentum'' $x$ of a vertex
$\rme^{x\varphi}$ by $n\alpha_+$ with integer~$n$, which corresponds
to changing $r$ in $\rme^{j(r,s)\varphi}$ by an \textit{even} integer.
It therefore follows that for each fixed $s\,{=}\,1,\dots,p$, the sum
\begin{equation*}
  \repF(s):=\bigoplus_{r\in\oZ}\rep{F}_{j(r,s)}
\end{equation*}
of Fock modules contains \textit{two} $\algW(p)$ modules, to be
denoted by $\repLambda(s)$ and $\repPi(s)$, where $\repLambda(s)$~is
the $\algW(p)$~representation generated from $\rme^{j(1,s)\varphi}$
(the highest-weight vector in $\rep{F}_{j(1,s)}$), while $\repPi(s)$
is the $\algW(p)$ representation generated from $\rme^{j(2,s)\varphi}$
(the highest-weight vector in $\rep{F}_{j(2,s)}$), see
Fig.~\ref{fig:LambdaPi}.
The dimensions of the corresponding highest-weight vectors are given
by
\begin{equation}
  \Delta_{\repLambda(s)}-\ffrac{c}{24}=\ffrac{(p-s)^2}{4p}-\ffrac{1}{24},
  \qquad
  \Delta_{\repPi(s)}-\ffrac{c}{24}=\ffrac{(2p-s)^2}{4p}-\ffrac{1}{24}.
\end{equation}

A somewhat more involved analysis shows that the corresponding kernel
of the screening $S_-$,
\begin{equation*}
  \repK(s):=\Ker S_-\!\!\bigm|_{\repF(s)},\quad s=1,\dots,p,
\end{equation*}
is precisely the direct sum
\begin{equation*}
  \repK(s)=\repLambda(s)\oplus\repPi(s).
\end{equation*}

\subsection{Extensions among the representations}\label{rep-exts}
We next describe the nontrivial extensions allowed by the $\algW(p)$
representations.  The category of representations of $\algW(p)$ in the
$(1,p)$ model decomposes into \textit{linkage classes} of
representations, which are full subcategories of the representation
category.\footnote{The term ``linkage class'' is borrowed from the
  theory of finite-dimensional Lie algebras. The linkage classes of an
  additive category $\catC$ are additive full subcategories $\catC_i$
  such that there are no (nonzero) morphisms between objects in two
  distinct linkage classes, every object of $\catC$ is a direct sum of
  objects of the linkage classes, and none of the $\catC_i$ can be
  split further in the same manner.}

\smallskip
 
The representation category of the algebra $\algW(p)$ associated with
the $(1,p)$ model has~$p\,{+}\,1$ linkage classes; we denote them
as~$\lc$, $\lc'$, and~$\lc(s)$ with $1\,{\leq}\, s\,{\leq}\, p{-}1$.
The indecomposable representations in each linkage class are as
follows.  The classes $\lc$ and~$\lc'$ contain only a single
indecomposable (hence, irreducible) representation each,
$\repLambda(p)$ and~$\repPi(p)$ respectively.  For $1\,{\leq}\, s
\,{\leq}\,p{-}1$, the linkage class~$\lc(s)$ contains two irreducible
representations~$\repLambda(s)$ and~$\repPi(p{-}s)$, as well as the
following set of other indecomposable representations:
\begin{equation*}
  \rep{N}_0^{\pm}(s),\quad
  \rep{N}_0(s),\quad
  \rep{N}_1^{\pm}(s),\quad
  \rep{N}_1(s),\quad
  \rep{R}_0(s),\quad
  \rep{R}_1(s).
\end{equation*}
There are nontrivial extensions
\begin{align*}
  &0\to\repLambda(s)\to\rep{N}_0^{\pm}(s)\to\repPi(p{-}s)\to0,\\
  &0\to\repPi(p{-}s)\to\rep{N}_1^{\pm}(s)\to\repLambda(s)\to0,
\end{align*}
and in addition,
\begin{align*}
  &0\to\repLambda(s)\to\rep{N}_0(s)\to
  \repPi(p{-}s)\,{\oplus}\,\repPi(p{-}s)\to0,\\*
  &0\to\repPi(p{-}s)\to\rep{N}_1(s)\to
  \repLambda(s)\,{\oplus}\,\repLambda(s)\to0.
\end{align*}
We note that $L_0$ is diagonalizable in each of these representations.
The ``logarithmic'' modules (those with a nondiagonalizable action
of~$L_0$) appear in the extensions
\begin{equation*}
  0\to\rep{N}_0(s)\to\rep{R}_0(s)\to\repLambda(s)\to0,\quad
  0\to\rep{N}_1(s)\to\rep{R}_1(s)\to\repPi(p{-}s)\to0.
\end{equation*}
It follows that
$\rep{N}_0^+(s)\,{\cap}\,\rep{N}_0^-(s)\,{=}\,\repLambda(s)$ and
$\rep{N}_1^+(s)\,{\cap}\,\rep{N}_1^-(s)\,{=}\,\repPi(p{-}s)$.  Thus we
have towers of indecomposable representations given by
\begin{gather*}
  \rep{R}_0(s)\supset\rep{N}_0(s)\supset\rep{N}_0^{\pm}(s)
    \supset\repLambda(s),\qquad
  \rep{R}_1(s)\supset\rep{N}_1(s)\supset\rep{N}_1^{\pm}(s)
   \supset\repPi(p{-}s)
\end{gather*}
for each $s\,{=}\,1,\dots,p\,{-}\,1$.  The detailed structure of these
representations will be considered elsewhere (see more comments in the
Conclusions, however).


\begin{example}
  For $p\,{=}\,2$, the four irreducible representations are
  $\Vi\,{=}\,\repLambda(2)$, $\Viii\,{=}\,\repPi(2)$,
  $\rep{V}_0\,{=}\,\repLambda(1)$, and $\rep{V}_1\,{=}\,\repPi(1)$.
  The ``logarithmic'' modules are $\rep{R}_0\,{=}\,\rep{R}_0(1)$
  and~$\rep{R}_1\,{=}\,\rep{R}_1(1)$~\cite{GK1}.  In addition, there
  are six other indecomposable representations $\rep{N}^\pm_0$,
  $\rep{N}^{}_0$, $\rep{N}^\pm_1$, and $\rep{N}^{}_1$.
\end{example}

\section{Modular transformations  of the $\algW(p)$ characters}
In this section, we evaluate the characters of the $\algW(p)$
representations introduced above and find their modular transformation
properties.

\subsection{Calculation of the $\algW(p)$ characters}\label{chars}
The route from representations to fusion starts with the characters of
$\algW(p)$ representations.  We write $\chi^{\repXi}_{s,p}$ with
$\repXi\,{\in}\,\{\repLambda,\repPi\}$ for the character of
$\repXi(s)$ in the~$(1,p)$ model,
\begin{equation*}
  \chi^{\repXi}_{s,p}(q)
  =\langle q^{L_0-\frac{c}{24}}\rangle^{}_{\!\repXi(s)}.
\end{equation*}


\begin{Prop}The $\algW(p)$ characters are given by
  \begin{equation}\label{eq:characters}
    \begin{aligned}
      \chi^{\repLambda}_{s,p}(q)&=\mfrac{1}{\eta(q)}
      \Bigl(\ffrac{s}{p}\,\theta_{p{-}s,p}(q)
      + 2\,\theta'_{p{-}s,p}(q)\Bigr),\\[4pt]
      \chi^{\repPi}_{s,p}(q)&=
      \mfrac{1}{\eta(q)}
      \Bigl(\ffrac{s}{p}\,\theta_{s,p}(q) - 2\,\theta'_{s,p}(q)\Bigr),
    \end{aligned}\qquad
    1\leq s\leq p.
  \end{equation}
\end{Prop}
\noindent
Here, we use the eta function
\begin{align*}
  \eta(q)&=q^{\frac{1}{24}} \prod_{n=1}^{\infty} (1-q^n)
  \\
  \intertext{and the theta functions} 
  \theta_{s,p}(q,z)&=\sum_{j\in\oZ + \frac{s}{2p}} q^{p j^2} z^j,
  \quad |q|<1,~z\in\oC\,,
\end{align*}
and set $\theta_{s,p}(q)\,{:=}\,\theta_{s,p}(q,1)$ and
$\theta'_{s,p}(q)\,{:=}\,z\frac{\dd}{\dd
  z}\theta_{s,p}(q,z)\!\!\bigm|_{z=1}$.

\begin{proof} Formulas~\eqref{eq:characters} (which also have been
  suggested in~\cite{flohr-0111228}) can be derived by standard
  calculations, which we outline here for completeness.
  
  The characters of $\repLambda(s)$ and $\repPi(s)$ are found by
  summing the characters of the kernels of $S_-$ on the appropriate
  Fock modules,
  \begin{align*}
    \chi^{\repLambda}_{s,p}&=\chr{\rep{K}(1,s)} +
    2\sum_{a\geq1}\chr{\rep{K}(2a{+}1,s)},\\
    \chi^{\repPi}_{s,p}&=2\sum_{a\geq1}\chr{\rep{K}(2a,s)},
  \end{align*}
  where
  \begin{equation*}
    \rep{K}(r,s):=\Ker S_-\!\!\bigm|_{\rep{F}_{j(r,s)}}.
  \end{equation*}
  The character of $\rep{K}(r,s)$, in turn, is easily calculated from
  a ``half'' of the complex~\eqref{Felder}, i.e., from the one-sided
  resolution, as either the kernel or the image of the corresponding
  differential, which amounts to taking the alternating sum of
  characters of the modules in the left or right part of the complex.
  A standard calculation (with some care to be taken in rearranging
  double sums) then gives the formulas in the proposition.
\end{proof}

\subsection{$S$ and $T$ transformations of the
  characters}\label{sec:fancy-S}
With the characters of $\repLambda(s)$ and $\repPi(s)$ expressed
through theta functions, it is straightforward to find their modular
properties.  We resort to the standard abuse of notation by writing
$\theta_{s,p}(\tau)$ for $\theta_{s,p}(\rme^{2\rmi\pi\tau})$, for
$\tau$ in the upper complex half-plane
$\upperH$.\enlargethispage{1.1\baselineskip}


\begin{Prop} Under the $S$ transformation of~$\tau$, the $\algW(p)$
  characters transform as
  \begin{multline*}
    \chi_{s,p}^{\repLambda}(-\ffrac{1}{\tau})
    =\ffrac{1}{\sqrt{2p}}\biggl( \ffrac{s}{p}\,\Bigl[
    \chi_{p,p}^{\repLambda}(\tau) +
    (-1)^{p-s}\chi_{p,p}^{\repPi}(\tau)\\*
    + 2\sum_{\ell=1}^{p-1}\cos\bigl(\pi\ffrac{\ell(p{-}s)}{p}\bigr)
    \bigl(\chi_{p-\ell,p}^{\repLambda}(\tau) +
    \chi_{\ell,p}^{\repPi}(\tau)\bigr)
    \Bigr]\\*[-2pt]
    \smash[b]{ - 2 \rmi\tau
      \sum_{\ell=1}^{p-1}\sin\bigl(\pi\ffrac{\ell(p{-}s)}{p}\bigr)
      \bigl(\ffrac{\ell}{p}\,\chi_{p-\ell,p}^{\repLambda}(\tau) -
      \ffrac{p{-}\ell}{p}\,\chi_{\ell,p}^{\repPi}(\tau)\bigr) \biggr)}
  \end{multline*}
and
  \begin{multline*}
    \chi_{s,p}^{\repPi}(-\ffrac{1}{\tau}) =\ffrac{1}{\sqrt{2p}}\biggl(
    \ffrac{s}{p}\Bigl[
    \chi_{p,p}^{\repLambda}(\tau) + (-1)^{s}\chi_{p,p}^{\repPi}(\tau)\\*
    + 2\sum_{\ell=1}^{p-1}\cos\bigl(\pi\ffrac{\ell s}{p}\bigr)
    \bigl(\chi_{p-\ell,p}^{\repLambda}(\tau) +
    \chi_{\ell,p}^{\repPi}(\tau)\bigr)
    \Bigr]\\*[-2pt]    
      + 2 \rmi\tau \smash[b]{\sum_{\ell=1}^{p-1}}
    \sin\bigl(\pi\ffrac{\ell s}{p}\bigr)
    \bigl(\ffrac{\ell}{p}\chi_{p-\ell,p}^{\repLambda}(\tau) -
    \ffrac{p-\ell}{p}\chi_{\ell,p}^{\repPi}(\tau)\bigr) \biggr)
  \end{multline*}
  $($with $\rmi\,{=}\,\sqrt{-1})$.
\end{Prop}

The functions $\theta_{s,p}$ and $\theta'_{s,p}$ are modular forms of
different weights ($\half$ and $\frac{3}{2}$ respectively) and do not
therefore mix in modular transformations.  In contrast, the characters
are linear combinations of modular forms of weights $0$ and~$1$ and
hence involve explicit occurrences of $\tau$ in their $S$
transformation.
\begin{proof}
  The formulas in the proposition are shown by directly applying the
  well-known relations
  \begin{align*}
    \theta_{s,p}(-\ffrac{1}{\tau}) &=\sqrt{\ffrac{-\rmi\tau}{2p}}
    \Bigl(\theta_{0,p}(\tau) + (-1)^s\theta_{p,p}(\tau) +
    2\sum_{\ell=1}^{p-1}\cos\bigl(\pi\ffrac{\ell s}{p}\bigr)
    \,\theta_{\ell,p}(\tau)
    \Bigr),\\
    \smash[b]{\theta'_{s,p}(-\ffrac{1}{\tau})} &=
    \smash[b]{-2\rmi\tau\sqrt{\ffrac{-\rmi\tau}{2p}}
      \sum_{\ell=1}^{p-1}\sin\bigl(\pi\ffrac{\ell s}{p}\bigr)
      \,\theta'_{\ell,p}(\tau).}
  \end{align*}
\end{proof}  

\subsubsection{The $\fancyS_p(\tau)$ matrix} 
We now write the $S$ transformation in a matrix form.  To this end, we
order the representations as
\begin{equation}\label{rep-order}
  \repLambda(p),\; \repPi(p),\; \repLambda(1),\;
  \repPi(p{-}1),\;\dots\,,\;
  \repLambda(p{-}1),\;\repPi(1),
\end{equation}
and arrange the characters into a column vector $\bs{\chi}_p$,
\begin{equation*}
  \bs{\chi}_p^{t}=
  (\,\chi^{\repLambda}_{p,p},\,
  \chi^{\repPi}_{p,p},\,
  \chi^{\repLambda}_{1,p},\,
  \chi^{\repPi}_{p-1,p},\,
  \dots,\,
  \chi^{\repLambda}_{p-1,p},\,
  \chi^{\repPi}_{1,p}\,).
\end{equation*}
This order is chosen such that representations in the same linkage
class are placed next to each other; it is one of the ingredients that
make the block structure explicit in what follows.  The above
$S$ transformation formulas then become
\begin{gather}\label{fancyS-acts}
  \bs{\chi}_p(-\ffrac{1}{\tau})= \fancyS_p(\tau)\, \bs{\chi}_p(\tau),
\end{gather}
where $\fancyS_p(\tau)$ is most conveniently written using the
$2\,{\times}\,2$ block notation
\begin{gather}
  \fancyS_p(\tau)=
  \begin{pmatrix}
    A_{0,0}&A_{0,1}&\dots&A_{0,p-1}\\
    A_{1,0}&A_{1,1}&\dots&A_{1,p-1}\\
    \hdotsfor{4}\\
    A_{p-1,0}&A_{p-1,1}&\dots&A_{p-1,p-1}
  \end{pmatrix}\label{fancyS}
\end{gather}
with
\begin{alignat*}{2}
  A_{0,0}&= \ffrac{1}{\sqrt{2p}}\begin{pmatrix}
    1\;&1\\
    1\;&(-1)^p 
  \end{pmatrix}\!,&
  A_{0,j}&= \ffrac{2}{\sqrt{2p}}\begin{pmatrix}
    1\;&1\\
    (-1)^{p-j}\;&(-1)^{p-j}
  \end{pmatrix}\!,\\[4pt]
  A_{s,0}&=\ffrac{1}{\sqrt{2p}}
  \begin{pmatrix}
    \ffrac{s}{p}\;   &(-1)^{p+s}\ffrac{s}{p}\\[9pt]
    \ffrac{p{-}s}{p}\; &(-1)^{p+s}\ffrac{p{-}s}{p}
  \end{pmatrix}\!,
\end{alignat*}
and
\begin{multline*}
  A_{s,j}=\sqrt{\ffrac{2}{p}}\,
  (-1)^{p+j+s}\times{}\\*
  \begin{pmatrix}
    \ffrac{s}{p}\cos\pi\ffrac{sj}{p}
    -\rmi\tau\,\ffrac{p{-}j}{p}\sin\pi\ffrac{sj}{p}\;
    &\ffrac{s}{p}\cos\pi\ffrac{sj}{p}
    +\rmi\tau\,\ffrac{j}{p}\sin\pi\ffrac{sj}{p}\\[10pt]
    \ffrac{p{-}s}{p}\cos\pi\ffrac{sj}{p}
    +\rmi\tau\,\ffrac{p{-}j}{p}\sin\pi\ffrac{sj}{p}\;
    &\ffrac{p{-}s}{p}\cos\pi\ffrac{sj}{p}
    -\rmi\tau\,\ffrac{j}{p}\sin\pi\ffrac{sj}{p}
  \end{pmatrix}
\end{multline*}
for $1\,{\leq}\, s,j\,{\leq}\, p{-}1$.

\subsubsection{The $\fancyT_p$ matrix} 
We next find the $T$ transformation of the $\algW(p)$ characters.  For
the vector $\bs{\chi}_p$ introduced above, we have
\begin{gather}\label{fancyT-acts}
  \bs{\chi}_p(\tau{+}1)= \fancyT_p\,\bs{\chi}_p(\tau),
\end{gather}
where $\fancyT_p$ is a block-diagonal matrix that can be compactly
written as a direct sum of $2\,{\times}\,2$ blocks,
\begin{gather}\label{fancyT}
  \fancyT_p=T_{0}\oplus T_{1}\oplus\dots\oplus T_{p-1}\\
  \intertext{with} T_{0}=
  \begin{pmatrix}
    \rme^{-\rmi\frac{\pi}{12}}&0\\
    0&\rme^{\rmi\pi(\frac{p}{2}-\frac{1}{12})}
  \end{pmatrix}\!,
  \qquad T_{s}=\rme^{\rmi\pi(\frac{(p-s)^2}{2p}-\frac{1}{12})}
  \,{\boldsymbol{1}}^{}_{2\times2}, \quad s=1,\dots,p{-}1.
  \label{T-blocks}
\end{gather}


Starting from the $\algW(p)$ algebra in $(1,p)$ models, we have thus
arrived at the $\fancyS_p(\tau)$ and $\fancyT_p$ matrices that
implement modular transformations on the characters.  Problem
Q\ref{which-algebra} in the Introduction has thus been solved.  With
the resulting $\fancyS_p(\tau)$ involving a dependence on~$\tau$, we
next face problem~Q\ref{involve-tau}, to be addressed in the next
section.

\section{A finite-dimensional $\SLiiZ$ representation from
  characters}\label{sec:char-to-S}
\subsection{Matrix automorphy factors}\label{automorphy}
The modular group action on characters generated
by~\eqref{fancyS-acts} and~\eqref{fancyT-acts} fits the following
general scheme.  It is well known (or easily checked) that
\begin{gather}\label{act-vector}
  (\gamma\,{\cdot}\, f)(\tau,\nu):=j(\gamma;\tau,\nu)\,
  f(\gamma\tau,\gamma\nu),
\end{gather}
with $\,j(\gamma;\tau,\nu)$ an $n\,{\times}\,n$\,-matrix satisfying the
cocycle condition
\begin{gather}\label{eq:cocycle}
  j(\gamma\gamma';\tau,\nu)= j(\gamma';\tau,\nu)\,
  j(\gamma;\gamma'\tau,\gamma'\nu),\qquad j(\one;\tau,\nu)=\one_{n\times n},
\end{gather}
furnishes an action (actually, an \textit{anti}representation) of the
modular group $\SLiiZ$ on the space of
functions~$f{:}\;\upperH\,{\times}\,\oC\,{\to}\,\oC^{n}$.  We use the
standard $\SLiiZ$ action on~$\upperH\,{\times}\,\oC$\,,
\begin{equation*}
  \gamma=
  \begin{pmatrix}
    a\;&b\\
    c\;&d
  \end{pmatrix}
  \!{:}\quad (\tau,\nu)\,\mapsto\,
  (\gamma\tau,\gamma\nu):=
  \Bigl(\ffrac{a\tau+b}{c\tau+d}, \ffrac{\nu}{c\tau+d}\Bigr)
\end{equation*}
(the notation $\gamma\nu$ is somewhat loose, because this action
depends on~$\tau$).  The matrix $j(\gamma;\tau,\nu)$ is called the
(matrix) \textit{automorphy factor}.

An example of a \textit{scalar} automorphy factor is given by the
following classic result in the theory of theta
functions~\cite{Mumford}: the Jacobi theta function
$\vartheta(\tau,\nu)$ is invariant under the action of
$\Gamma_{\!1,2}\,{\subset}\,\SLiiZ$ (the subgroup of $\SLiiZ$ matrices
$\gamma\,{=}\,\smatrix{a}{b}{c}{d}$ with $ab$ and $cd$ even) on
functions $f{:}\;\upperH\,{\times}\,\oC\,{\to}\,\oC\,$ given by
\begin{equation}\label{act-scalar}
  (\gamma\,{\cdot}\, f)(\tau,\nu)=j(\smatrix{a}{b}{c}{d};\tau,\nu)
  \,f(\gamma\tau, \gamma\nu)
\end{equation}
with the automorphy factor
\begin{equation*}
  j(\smatrix{a}{b}{c}{d};\tau,\nu)=
  \zeta_{c,d}^{-1}\,(c\tau{+}d)^{-\half}_{}\,
  \rme^{-\rmi\pi\frac{c\nu^2}{c\tau + d}}_{},
\end{equation*}
where $\zeta_{c,d}$ is an eighth root of unity (see~\cite{Mumford};
its definition, which is far from trivial, ensures the cocycle
condition for~$j$).

\subsection{Constructing a finite-dimensional $\SLiiZ$ representation}
\label{representation}
The $\algW(p)$ characters that we study here do not involve the $\nu$
dependence.  Because $S\,{=}\,\smatrix{0}{-1}{1}{0}$ and
$T\,{=}\,\smash[t]{\smatrix{1}{1}{0}{1}}$ generate $\SLiiZ$,
Eqs.~\eqref{fancyS-acts} and~\eqref{fancyT-acts} uniquely determine a
$2p\,{\times}\,2p$ matrix $\bs{J}_p(\gamma,\tau)$ such that 
\begin{equation*}
  \bs{\chi}_p(\gamma\tau)=\bs{J}_p(\gamma,\tau)\,\bs{\chi}_p(\tau)
\end{equation*}
for all $\gamma\,{\in}\,\SLiiZ$.  It then follows that $\bs{J}_p$
satisfies the condition
\begin{equation}\label{J-condition}
  \bs{J}_p(\gamma\,\gamma',\tau)=
  \bs{J}_p(\gamma,\gamma'\tau)\,\bs{J}_p(\gamma',\tau),
  \qquad \gamma,\gamma'\in\SLiiZ.
\end{equation}

Given this $\SLiiZ$ action, we now seek an $\SLiiZ$ action on
$\bs{\chi}_p{:}\;\upperH\,{\to}\,\oC^{2p}$ with a $2p\,{\times}\,2p$
matrix automorphy factor $j_p$,
\begin{equation*}
  \gamma\cdot\bs{\chi}_p(\tau)=j_p(\gamma,\tau)\,\bs{\chi}_p(\gamma\tau)
  =j_p(\gamma,\tau)\,\bs{J}_p(\gamma,\tau)\,\bs{\chi}_p(\tau),
\end{equation*}
such that
\begin{equation}\label{pi-rep}
  \rho(\gamma):=j_p(\gamma,\tau)\,\bs{J}_p(\gamma,\tau)
\end{equation}
is a finite-dimensional \textit{representation} of $\SLiiZ$ (in
particular, the left-hand side must be independent of~$\tau$).  This
condition is reformulated as the condition that $\rho$ and $j_p$
``strongly'' commute, i.e., that
\begin{gather}\label{strong-commute}
  \rho(\gamma)\,j_p(\gamma', \tau)
  =j_p(\gamma', \tau)\,\rho(\gamma), \qquad
  \gamma,\gamma'\in\SLiiZ.
\end{gather}

It is easy to verify that for a given
$\bs{J}_p({}\,{\cdot}\,{},{}\,{\cdot}\,{})$, each $j_p$ that satisfies
both the commutation property~\eqref{strong-commute} (with $\rho$
defined by~\eqref{pi-rep}) and the cocycle condition~\eqref{eq:cocycle}
provides a (finite-dimensional) $\SLiiZ$ \textit{representation}
$\rho$.  Indeed,
\begin{multline*}
  \quad\rho(\gamma\gamma')=j_p(\gamma\gamma',\tau)\,
  \bs{J}_p(\gamma\gamma',\tau)
  =j_p(\gamma',\tau)\,\rho(\gamma)\,\bs{J}_p(\gamma',\tau)\\*
  =\rho(\gamma)\,j_p(\gamma',\tau)\,\bs{J}_p(\gamma',\tau)
  =\rho(\gamma)\,\rho(\gamma').\qquad
\end{multline*}

\subsection{$\SLiiZ$ representation in $(1,p)$
  models}\label{representation1p} 
We now find a matrix automorphy factor $j_p$ that ``converts'' the
action in~\eqref{fancyS-acts}\,--\,\eqref{fancyT} into a
\textit{representation}.  As noted above,
$\bs{J}_p({}\,{\cdot}\,{},\tau)$ is uniquely determined on all of
$\SLiiZ$ by Eqs.~\eqref{J-condition} from
$\bs{J}_p(T,\tau)\,{=}\,\fancyT_p(\tau)$ and
$\bs{J}_p(S,\tau)\,{=}\,\fancyS_p(\tau)$.  With $\fancyS_p(\tau)$ and
$\fancyT_p(\tau)\,{=}\,\fancyT_p$ given by~\eqref{fancyS}
and~\eqref{fancyT}, we define the automorphy
factor~$j_p({}\,{\cdot}\,{},{}\,{\cdot}\,{})$ as a block-diagonal
matrix consisting of $2\,{\times}\,2$ blocks that we compactly write
as
\begin{equation}\label{j-gen}
  j_p(\gamma,\tau)=
  \boldsymbol{1}_{2\times2}
  \oplus B_1(\gamma,\tau)
  \oplus\,\cdots\,\oplus B_{p-1}(\gamma,\tau),
\end{equation}
where for $\gamma\,{=}\,S$,
\begin{equation}\label{B(S)}
  B_s(S,\tau)=
  \begin{pmatrix}
    \ffrac{s}{p}+\rmi\,\ffrac{p{-}s}{\tau p}\;
    &\ffrac{s}{p}-\rmi\,\ffrac{s}{\tau p}\\[10pt]
    \ffrac{p{-}s}{p}-\rmi\,\ffrac{p{-}s}{\tau p}\;
    &\ffrac{p{-}s}{p}+\rmi\,\ffrac{s}{\tau p}
  \end{pmatrix}\!,
  \quad s=1,\dots,p\,{-}\,1,
\end{equation}
and for $\gamma\,{=}\,T$,
\begin{equation}\label{B(T)}
  B_s(T,\tau)=
  \begin{pmatrix}
    \ffrac{s}{p}+t\,\ffrac{p{-}s}{p}\;
    &\ffrac{s}{p}-t\,\ffrac{s}{p}\\[8pt]
    \ffrac{p{-}s}{p}-t\,\ffrac{p{-}s}{p}\;
    &\ffrac{p{-}s}{p}+t\,\ffrac{s}{p}
  \end{pmatrix}\!,
  \quad s=1,\dots,p\,{-}\,1,
\end{equation}
with $t^3\,{=}\,{-}\rmi$ (we can set $t\,{=}\,\rmi$).  The structure
in~\eqref{B(S)} is easily discernible by subjecting all matrices to
the similarity transformation that relates the basis of characters to
the basis provided by $\theta_{s,p}$ and $\theta_{s,p}'$.  The
automorphy factor is then diagonalized, as shown explicitly in the
proof of the next proposition.


\begin{Prop}\label{prop:cocycle}
  The matrix automorphy factor defined
  in~\eqref{j-gen}\,--\,\eqref{B(T)} satisfies the cocycle
  condition~\eqref{eq:cocycle}
\end{Prop}
\begin{proof}
  The proof amounts to a direct verification of the formulas
  $(ST)^3\,{=}\,(TS)^3\!=S^2$ reformulated for
  $j_p(\gamma,\tau)$.  That is, in proving that
  $j_p(S^2,\tau)\,{=}\,j_p((S T)^3,\tau)$, we have, in accordance
  with~\eqref{eq:cocycle},
  \begin{equation}\label{prove-this}
    \begin{split}
      j_p(S^2,\tau)&=j_p(S,\tau)\,j_p(S,-\ffrac{1}{\tau}),\\
      j_p((S T)^3,\tau)&=j_p(S T,\tau)\,j_p(S T,\ffrac{-1}{\tau{+}1})
      \,j_p(S T,\ffrac{-\tau{-}1}{\tau}),
    \end{split}
  \end{equation}
  where in turn, $j_p(S T,\tau)\,{=}\,j_p(T,\tau)j_p(S,\tau{+}1)$.
  The calculation reduces to a separate computation for each of the
  $2\,{\times}\,2$ blocks given above; further, each block can be
  diagonalized~as
  \begin{equation*}
    B_s(\gamma,\tau)\,{=}\,L_s
    \begin{pmatrix}1&0\\
      0\;&\zeta(\gamma)\alpha(\gamma,\tau)
    \end{pmatrix}
    L_s^{-1},
  \end{equation*}
  where $\zeta(\gamma)$ is the character of~$\SLiiZ$ defined by the
  relations
  \begin{equation}\label{eq:sl2zchar}
    \zeta(S)=\rmi,\qquad\zeta(T)=t,\qquad t^3=-\rmi,
  \end{equation}
  and
  \begin{equation*}
    \smash[t]{\alpha(\smatrix{a}{b}{c}{d},\tau)=\ffrac{1}{c\tau+d}}
  \end{equation*}
  is already an automorphy factor~\cite{Mumford}.
  Equations~\eqref{eq:sl2zchar} immediately imply that
  $\zeta(S^2)\,{=}\,\zeta((ST)^3)$, and Eqs.~\eqref{prove-this} are
  therefore proved.
\end{proof}

With this $j_p$, we evaluate
$\pureS(p)\,{=}\,j_p(S,\tau)\fancyS_p(\tau)$ as
\begin{equation}\label{pureS}
  \pureS(p)=j_p(S,\tau)\,\fancyS_p(\tau)
  =\fancyS_p(\rmi).
\end{equation}
That is, $\pureS(p)$ has a block form similar to that of $\fancyS_p$
in Sec.~\ref{sec:fancy-S}, with the $2\,{\times}\,2$ blocks $S_{i,j}$
given by $S_{0,0}\,{=}\,A_{0,0}$, $S_{0,j}\,{=}\,A_{0,j}$,
$S_{s,0}\,{=}\,A_{s,0}$, and
\begin{equation*}
  S_{s,j}=\sqrt{\ffrac{2}{p}}
  (-1)^{p+j+s}
  \begin{pmatrix}
    \ffrac{s}{p}\cos\pi\ffrac{sj}{p}
    +\ffrac{p{-}j}{p}\sin\pi\ffrac{sj}{p}\;
    &\ffrac{s}{p}\cos\pi\ffrac{sj}{p}
    -\ffrac{j}{p}\sin\pi\ffrac{sj}{p}\\[10pt]
    \ffrac{p{-}s}{p}\cos\pi\ffrac{sj}{p}
    -\ffrac{p{-}j}{p}\sin\pi\ffrac{sj}{p}\;
    &\ffrac{p{-}s}{p}\cos\pi\ffrac{sj}{p}
    +\ffrac{j}{p}\sin\pi\ffrac{sj}{p}
  \end{pmatrix}\!.
\end{equation*}


Similarly,
\begin{equation*}
  \pureT(p)=j_p(T,\tau)\,\fancyT_p
\end{equation*}
(where as we have seen, $j_p(T,\tau)$ is actually independent
of~$\tau$).  We do not write the blocks of $\pureT(p)$ explicitly
because they are simply given by multiplication of the blocks
in~\eqref{B(T)} with matrices~\eqref{T-blocks}.


\begin{Prop}
  The matrices $\pureS(p)$ and $\pureT(p)$ generate a
  finite-dimensional representation of $\SLiiZ$.
\end{Prop}
\begin{proof}
  The proof consists in verifying~\eqref{strong-commute} for
  $(\gamma,\gamma')$ being any of the pairs $(S,T)$, $(T,S)$, $(S,S)$,
  and $(T,T)$, which is straightforward.  Together with the cocycle
  condition, this then implies that
  $(\pureS(p))^2\,{=}\,(\pureT(p)\pureS(p))^3\,{=}\,
  (\pureS(p)\pureT(p))^3$.
\end{proof}

The above construction of the numeric ($\tau$-independent) matrix
$\pureS(p)$ representing $S\,{\in}\,\SLiiZ$ solves
problem~Q\ref{involve-tau} in the Introduction.

\subsection{Some properties of the $\pureS(p)$
  matrix}\label{properties}
The vacuum representation $\repLambda(1)$ is the third in the order of
representations chosen in~\eqref{rep-order}. This distinguishes the
third row of the $\pureS$ matrix; we let
$\sigma_\Omega(p)\,{\equiv}\,\sigma_\Omega$ denote this distinguished
row of~$\pureS(p)$.  Explicitly, $\sigma_\Omega(p)$ is given by
\begin{multline}\label{sigma-sharp} \sigma_\Omega(p)=
  (-1)^p\ffrac{\sqrt{2}}{p\sqrt{p}}\, \Bigl( \ffrac{(-1)^p}{2},
  -\ffrac{1}{2},\\
  \shoveleft{\qquad \cos\ffrac{\pi}{p} +(p{-}1)\sin\ffrac{\pi}{p},\;
    \cos\ffrac{\pi}{p}
    -\sin\ffrac{\pi}{p},}\\
  \shoveleft{\quad\qquad -\cos\ffrac{2\pi}{p}
    -(p{-}2)\sin\ffrac{2\pi}{p},\; -\cos\ffrac{2\pi}{p}
    +2\sin\ffrac{2\pi}{p},}\\
  \dots,\\
  \qquad (-1)^{j+1} \bigl(\cos\ffrac{j\pi}{p}
  +(p{-}j)\sin\ffrac{j\pi}{p}\bigr),\; (-1)^{j+1}
  \bigl(\cos\ffrac{j\pi}{p}
  -j\sin\ffrac{j\pi}{p}\bigr),\\
  \dots,\\
  (-1)^{p} \bigl(\cos\ffrac{(p{-}1)\pi}{p}
  +\sin\ffrac{j\pi}{p}\bigr),\; (-1)^{p}
  \bigl(\cos\ffrac{(p{-}1)\pi}{p}
  -(p{-}1)\sin\ffrac{(p-1)\pi}{p}\bigr) \Bigr).
\end{multline}

Next, it follows from~\eqref{fancyS-acts} that
$(\pureS(p))^2\bs{\chi}_p(\rmi)\!=\!\bs{\chi}_p(\rmi)$.  In fact, we
have the following result.


\begin{Prop}\label{prop:S2}
  \begin{equation}\label{eq:S2}
    (\pureS(p))^2=\one_{2p \times 2p}.
  \end{equation}
\end{Prop}

\begin{proof}  
  Indeed, we evaluate $(\pureS(p))^2$ as
  \begin{multline*}
    \rho(S)\,\rho(S)\stackrel{\mbox{\tiny\eqref{pi-rep}}}{=}
     \rho(S)\,j_p(S,\tau)\bs{J}_p(S,\tau)
     \stackrel{\mbox{\tiny\eqref{strong-commute}}}{=}
    j_p(S,\tau)\,\rho(S)\,\bs{J}_p(S,\tau)
    \stackrel{\mbox{\tiny\eqref{pi-rep}}}{=}\\*
    j_p(S,\tau)\,j_p(S,S\tau)\,\bs{J}_p(S,S\tau)\,\bs{J}_p(S,\tau)
    \stackrel{\mbox{\tiny\eqref{J-condition}}}{=}
    j_p(S,\tau)\,j_p(S,S\tau)\,\bs{J}_p(S^2,\tau).
  \end{multline*}
  Next, $\bs{J}_p(S^2,\tau)\,{=}\,\one_{2p \times 2p}$ because
  $S^2\tau\,{=}\,\tau$.  Finally we have $j_p(S,\tau)\,j_p(S,S\tau)\,{=}
    $\linebreak[0]$%
   \one_{2p\times 2p}$, which is obtained by a direct calculation
  similar to the one in the proof of Prop.~\ref{prop:cocycle}.
  Equation~\eqref{eq:S2} thus follows.
\end{proof}


\begin{Rem}
  With the explicit form of $\pureS(p)$ given above,
  Prop.~\ref{prop:S2} can also be shown directly, which gives a good
  illustration of a typical calculation with the matrices encountered
  throughout this paper.  Writing $\chargeC\,{=}\,(\pureS(p))^2$ in
  the $2\,{\times}\,2$-block form
  \begin{equation*}
    \chargeC=
    \begin{pmatrix}
      C_{0,0}&C_{0,1}&\dots&C_{0,p-1}\\
      C_{1,0}&C_{1,1}&\dots&C_{1,p-1}\\
      \hdotsfor{4}\\
      C_{p-1,0}&C_{p-1,1}&\dots&C_{p-1,p-1}
    \end{pmatrix}\!,
  \end{equation*}
  we concentrate on the more involved blocks $C_{s,j}$ with
  $0\,{<}\,s,j\,{<}\,p$.  Assuming that $p$ is odd for brevity (in
  order to avoid extra sign factors) we find that
  \begin{multline*} \!\!
    C_{s,j}= \ffrac{2}{p^2}\times{}\\*
    \ \ \mbox{\small$\displaystyle
      \begin{pmatrix}
        \mbox{}\kern-4pt
        \begin{array}[b]{l}      
          s\!\sum\limits_{\ell=[s + j]_2}^{p - 1}\!
          \cos\pi\frac{\ell(p - j)}{p}\cos\pi\frac{\ell(p - s)}{p}\\
          \quad{}+ (p{-}j)\!\sum\limits_{\ell=1}^{p - 1}\!
          \sin\pi\frac{\ell(p - j)}{p}\sin\pi\frac{\ell(p - s)}{p}
        \end{array}\rule[-24pt]{0pt}{20pt}
        & \mbox{}
        \begin{array}[b]{l}
          s\!\sum\limits_{\ell=[s + j]_2}^{p - 1}\!
          \cos\pi\frac{\ell(p - j)}{p}\cos\pi\frac{\ell(p - s)}{p}\\
          \quad{}- j\sum\limits_{\ell=1}^{p - 1}
          \sin\pi\frac{\ell(p - j)}{p}\sin\pi\frac{\ell(p - s)}{p}
        \end{array}\kern-6pt\\[-2pt]
        \mbox{}\kern-4pt
        \begin{array}[b]{l}
          (p{-}s)\!\sum\limits_{\ell=[s + j]_2}^{p - 1}\!
          \cos\pi\frac{\ell(p - j)}{p}\cos\pi\frac{\ell(p - s)}{p}\\
          \qquad{}- (p{-}j)\!\sum\limits_{\ell=1}^{p - 1}\!
          \sin\pi\frac{\ell(p - j)}{p}\sin\pi\frac{\ell(p - s)}{p}
        \end{array}
        & \mbox{}
        \begin{array}[b]{l}
          (p{-}s)\!\sum\limits_{\ell=[s + j]_2}^{p - 1}\!
          \cos\pi\frac{\ell(p - j)}{p}\cos\pi\frac{\ell(p - s)}{p}\\
          \qquad{}+ j\sum\limits_{\ell=1}^{p - 1}
          \sin\pi\frac{\ell(p - j)}{p}\sin\pi\frac{\ell(p - s)}{p}
        \end{array}\kern-4pt
      \end{pmatrix}\!,$}
  \end{multline*}
  where $[a]_2\,{:=}\,a\,\mathrm{mod}\,2$.  Using elementary
  trigonometric rearrangements (expressing $\cos\alpha\,\sin\beta$
  through the sine and cosine of $\alpha{+}\beta$ and
  $\alpha{-}\beta$), we see that all entries in the matrices above
  vanish, with the exception of
  the diagonal entries of $C_{s,s}$, which (for
  $0\,{<}\,s\,{\leq}\,p$) are given by
  \begin{align*}
    \ffrac{2}{p^2}\, \Bigl(s \sum_{\ell=0}^{p - 1}
    \Bigl(\cos\pi\ffrac{\ell(p{-}s)}{p}\Bigr)^2 + (p{-}
    s)\sum_{\ell=1}^{p - 1}
    \Bigl(\sin\pi\ffrac{\ell(p{-}s)}{p}\Bigr)^2
    \Bigr)&=1.
  \end{align*}
  Together with similar (and in fact, simpler) calculations for the
  other blocks, this shows~\eqref{eq:S2}.
\end{Rem}

We also note that $\pureS(p)$ is not symmetric,
$\pureS(p)\,{\neq}\,\pureS(p)^{t}$.  It admits a different symmetry
\begin{equation}\label{vee}
  \pureS(p)^\vee=\pureS(p),
\end{equation}
where for a matrix $r\,{=}\,(r_{i,j})_{i,j=1,\dots,2p}$ with $i$ and $j$
considered modulo~$2p$, we define the involutive operation
\begin{equation*}
  (r^\vee)_{m,n}:=
  (-1)^{p (1 - \delta_{m,1} - \delta_{n,1}) + \floor{(m + n + 1)/2}
    + m n}\, r^{}_{2p - m + 3, 2p - n + 3}.
\end{equation*}
For example, with $r\,{=}\,(r_{ij})_{i,j=1,\dots,6}$, we have
\begin{equation*}
  r^\vee=
  \mbox{\small$\begin{pmatrix}
    -r_{2 2} & r_{2 1} & -r_{2 6} & -r_{2  5} & r_{2 4} & r_{2 3} \\
    r_{1 2} & -r_{1  1} & r_{1 6} & r_{1 5} & -r_{1 4} & -r_{1  3} \\
    -r_{6 2} & r_{6 1} & -r_{6 6} & -r_{6  5} & r_{6 4} & r_{6 3} \\
    -r_{5 2} & r_{5  1} & -r_{5 6} & -r_{5 5} & r_{5 4} & r_{5  3} \\
    r_{4 2} & -r_{4 1} & r_{4 6} & r_{4  5} & -r_{4 4} & -r_{4 3} \\
    r_{3 2} & -r_{3  1} & r_{3 6} & r_{3 5} & -r_{3 4} & -r_{3  3} 
  \end{pmatrix}\!.$}
\end{equation*}
The symmetry~\eqref{vee} originates in the existence of a simple
current, as we see below.


\begin{example}
  For $p=2$ and $p=3$, the $\pureS(p)$ matrices can be evaluated
  as
  \begin{align*}
    \pureS(2)&=\addtolength{\arraycolsep}{2pt}
    \begin{pmatrix}
      \frac{1}{2} & \frac{1} {2} & 1 & 1 \\[1pt]
      \frac{1}{2} & \frac{1}{2} & -1 & -1 \\[1pt]
      \frac{1}{4} & - \frac{1} {4} & \frac{1}{2} & - \frac{1}{2} \\[1pt]
      \frac{1} {4} & - \frac{1}{4} & - \frac{1}{2} & \frac{1}{2}
    \end{pmatrix}\!,
    \\[4pt]
    \pureS(3)&= \mbox{\standardfootnotesize$\displaystyle
      \begin{pmatrix}
        \frac{1}{{\sqrt{6}}} & \frac{1} {{\sqrt{6}}} &
        {\sqrt{\frac{2}{3}}} & { \sqrt{\frac{2}{3}}} &
        {\sqrt{\frac{2}{3}}} & {\sqrt{\frac{2}{3}}}
        \\[3pt]
        \frac{1}{{\sqrt{6}}} & - \frac{1} {{\sqrt{6}}} &
        {\sqrt{\frac{2}{3}}} & {\sqrt{\frac{2}{3}}} &
        -{\sqrt {\frac{2}{3}}} & -{\sqrt{\frac{2}{3}}} \\[4pt]
        \frac{1}{3{\sqrt{6}}} & \frac{1}{3{\sqrt{6}}} & \frac{-( 6 +
          {\sqrt{3}} ) }{9{\sqrt{2}}} & \frac{ 3 - {\sqrt{3}} }
        {9{\sqrt{2}}} & \frac{ 3 - {\sqrt{3}} }{9{\sqrt{2}}} &
        \frac{-( 6 + {\sqrt{3}} ) }
        {9{\sqrt{2}}} \\[3pt]
        \frac{{\sqrt{2/3}}}{3} & \frac{{\sqrt{2/3}}} {3} & \frac{
          {\sqrt{2}} ( 3 - {\sqrt{3}} ) } {9} & \frac{-( 3 +
          2{\sqrt{3}} ) }{9{\sqrt{2}}} & \frac{-( 3 + 2{\sqrt{3}} )
        }{9{\sqrt{2}}} & \frac{ {\sqrt{2}} ( 3 - {\sqrt{3}} ) }
        {9} \\[3pt]
        \frac{{\sqrt{2/3}}} {3} & \frac{-{\sqrt{2/3}}}{3} & \frac{
          {\sqrt{2}} ( 3 - {\sqrt{3}} ) } {9} & \frac{-( 3 +
          2{\sqrt{3}} ) }{9{\sqrt{2}}} & \frac{3 +
          2{\sqrt{3}}}{9{\sqrt{2}}} & \frac{ {\sqrt{2}}( {\sqrt{3}} -
          3 ) }
        {9} \\[3pt]
        \frac{1}{3{\sqrt{6}}} & \frac{-1} {3{\sqrt{6}}} & \frac{-( 6 +
          {\sqrt{3}} ) }{9{\sqrt{2}}} & \frac{ 3 - {\sqrt{3}} }
        {9{\sqrt{2}}} & \frac{{\sqrt{3}} - 3} {9{\sqrt{2}}} & \frac{6
          + {\sqrt{3}}} {9{\sqrt{2}}}
      \end{pmatrix}\!.$}
  \end{align*}
\end{example}

\section{Constructing the eigenmatrix $\eigenP$  and the
  fusion}\label{sec:buld-fusion}
Having extracted a finite-dimensional $\SLiiZ$ representation from the
$\SLiiZ$ action on characters, we now address problems
Q\ref{which-canonical} and~Q\ref{what-K} in the Introduction.  We use
the $\pureS(p)$ matrix found in the previous section in the
construction of the eigenmatrix $\eigenP$ of the fusion algebra.
{}From the eigenmatrix, we then find the fusion.
In~Sec.~\ref{sec:eigenP}, we first describe the role of the $\eigenP$
matrix in a commutative associative algebra in a slightly more general
setting than we actually need in $(1,p)$ models.  In
Sec.~\ref{sec:StoP}, we formulate the generalized Verlinde formula and
use it to find the eigenmatrix $\eigenP(p)$ in the $(1,p)$ model.  In
Sec.~\ref{sec:find-fusion}, we then obtain the fusion following the
recipe in Sec.~\ref{sec:eigenP}.

\subsection{Fusion constants from the eigenmatrix}\label{sec:eigenP}
A fusion algebra is a finite-dimen\-sional commutative associative
algebra~$\algF$ over $\oC$ with a unit element~$\boldsymbol{1}$,
together with a canonical basis~$\{X_I\}$,
$I\,{=}\,1,\dots,n\,{=}\,\dim_{\oC}\algF$
(containing~$\boldsymbol{1}$), such that the structure
constants~$N_{IJ}^{K}$ defined by
\begin{equation*}
  X_I\, X_J = \sum_{K=1}^n N_{IJ}^{K}\, X_K
\end{equation*}
are nonnegative integers.  As any finitely generated associative
algebra with a unit, $\algF$ is a vector-space sum of the
radical~$\radR$ and a semisimple algebra~\cite{Pierce}.  The algebra
contains a set of primitive idempotents satisfying
\begin{gather}\label{ee}
  e_A\,e_B=\delta_{A,B}\,e_B
\end{gather}
and 
\begin{equation}\label{eq:1-from-e}
  \smash[t]{\sum_{\substack{\text{all primitive}\\
        \text{idempotents}}}\!e_A=1.}
\end{equation}
The primitive idempotents characterize the semisimple quotient up to
Morita equivalence.  A \textit{commutative} associative algebra has a
basis given by the union of a basis in the radical and the primitive
idempotents~$e_A$.

The primitive idempotents can be classified by the dimensions $\nu_A$
of their images.  For the purposes of $(1,p)$ models, we only need to
consider the case where all $\nu_A\,{\leq}\,2$.\footnote{The fusion
  algebra for a general logarithmic conformal field theory can involve
  primitive idempotents with arbitrary~$\nu_A$.  We restrict our
  attention to the particular case where $\nu_A\,{\leq}\,2$ because of
  the lack of instructive examples of higher-``rank'' logarithmic
  theories; the definitions may need to be refined as further examples
  are worked out.  When the set of idempotents with $\nu_A\,{=}\,2$ is
  empty, we recover the semisimple case~\cite{Gannon} (we do not
  impose conditions \textbf{F2} and \textbf{F3} in~\cite{Gannon}
  because they imply semisimplicity of the fusion algebra).}  The
structure of the algebra~$\algF$ is then conveniently expressed by its
quiver
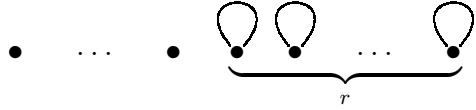
\begin{equation*}
  \rule{0pt}{24pt}
  {\xymatrix@=16pt{%
      *{\bullet}
      &{\dots}
      &*{\bullet}
      &}}\kern-8pt
  \underbrace{\xymatrix@=16pt{%
      *{\bullet}\ar@{-}@(lu,ru)[]
      &*{\bullet}\ar@{-}@(lu,ru)[]
      &\dots
      &*{\bullet}\ar@{-}@(lu,ru)[]
    }}_{r}
\end{equation*}
Here, the dots are in one-to-one correspondence with primitive
idempotents.  The quiver is disconnected because the algebra is
commutative.  A vertex $e_A$ has a self-link if $\nu_A\,{=}\,2$, and
has no links if $\nu_A\,{=}\,1$.  Each link can be associated with an
element in the radical, and moreover, these elements constitute a
\hbox{basis in the radical}.

We let $e_\alpha$ denote the primitive idempotents with
$\nu_\alpha\,{=}\,2$ and let $w_\alpha\,{\in}\,\radR$ be the
corresponding element, defined modulo a nonzero factor, represented by
the link of $e_\alpha$ with itself.
Then
\begin{equation}\label{ew}
  e_\alpha\,w_\beta=\delta_{\alpha,\beta}\,w_\beta.
\end{equation}
The other primitive idempotents, to be denoted by $e_a$, satisfy
\begin{equation}\label{ew0}
  e_a\,w_\beta=0.
\end{equation}
The elements $w_\alpha$ can be chosen such that they constitute a
basis in the radical and satisfy
\begin{equation}\label{ww}
  w_\alpha\,w_\beta=0.
\end{equation}

Let $Y_{\bullet}$ be the basis consisting of $e_a$, $e_\alpha$, and
$w_\alpha$; with $r$ introduced in the quiver above (as
$r\,{=}\,\dim_{\oC}\radR$), we have $a\,{=}\,1,\dots,n{-}2r$ and
$\alpha\,{=}\,n{-}2r{+}1,\dots,n{-}r$.  We order the elements in this
basis as
\begin{multline}\label{Y-basis}
  Y_1=e_1,\;\dots,\; Y_{n-2r}=e_{n-2r},\\*
  Y_{n-2r+1}=e_{n-2r+1},\; Y_{n-2r+2}=w_{n-2r+1},\\*[-2pt]
  \dots\dots,\\*[-4pt]
  Y_{n-1}=e_{n-r},\; Y_{n}=w_{n-r}.
\end{multline}
This ordering may seem inconvenient in that labeling of $w_\alpha$
starts with $w_{n-2r+1}$, but it is actually very useful in what
follows, because it makes the $2\,{\times}\,2$ block structure
explicit by placing each element $w_\alpha$ in the radical next to the
primitive idempotent~$e_\alpha$ that satisfies $e_\alpha
w_\alpha\,{=}\,w_\alpha$; the primitive idempotents that annihilate the
radical are given first.  It may be useful to rewrite~\eqref{Y-basis} as
\begin{equation*}
  Y_I=
  \begin{cases}
    e_I,             & I\,{=}\,1,2,\dots,n{-}2r,\\
    e_{(I+n+1)/2-r}, & I\,{=}\,n{-}2r{+}2i{+}1,\
    0\,{\leq}\,i\,{\leq}\,r{-}1,\\
    w_{(I+n)/2-r},   & I\,{=}\,n{-}2r{+}2i,\
    1\,{\leq}\,i\,{\leq}\,r.
  \end{cases}
\end{equation*} 

The multiplication table of $Y_\bullet$, Eqs.~\eqref{ee}\,--\,\eqref{ww},
defines an associative algebra.  But it does not define a fusion
algebra structure, because the latter involves a canonical basis.  The
canonical basis~$X_\bullet$ in~$\algF$ is specified by a nondegenerate
$n\,{\times}\, n$ matrix~$\eigenP$, called the eigenmatrix, that
contains a row entirely consisting of~$0$ ($r$ times) and~$1$
($n{-}r\,{\geq}\, r$ times).  We let $\pi_{\Omega}$ denote this row,
and order the columns of~$\eigenP$ in accordance with~\eqref{Y-basis},
such that
\begin{equation*}
  \pi_{\Omega}=\,
  (\,\underbrace{1,\dots,1}_{n-2r}\;
  \underbrace{1,0,1,0,\dots,1,0}_{2r}\,).
\end{equation*}

Elements of the canonical basis are given by
\begin{equation}\label{Y-to-X}
  X_I=\sum_{J=1}^n P_I^{\ J}Y_J
\end{equation}
and are therefore in one-to-one correspondence with the rows
of~$\eigenP$; permuting the rows of $\eigenP$ is equivalent to
relabeling the elements of the canonical basis.  The order of the
columns of $\eigenP$ is fixed by the assignments of $Y_\bullet$
in~\eqref{Y-basis}, i.e., by the order chosen for the elements of the
basis consisting of idempotents and elements in the radical, and is
therefore conventional. Each column corresponding to an element in
the radical (that is, containing zero in the intersection with the row
$\pi_\Omega$) is defined up to a factor, because $w_\alpha$ in the
radical cannot be canonically normalized.  In view
of~\eqref{eq:1-from-e}, it follows that $X_{\Omega}\,{=}\,\one$.

We now express the structure constants of the fusion algebra in the
canonical basis through a given eigenmatrix $\eigenP$.  We organize
the structure constants into matrices $N_I$ with the entries
\begin{gather*}
  (N_I)_J^{\ K}:=N_{IJ}^{K}.
\end{gather*}
Let $\pi_I\,{=}\,(P_I^{\ 1},\dots,P_{I}^{\ n})$ be the $I$th row
of~$\eigenP$.  For each $I\,{=}\,1,\dots,n$, we define the
$n\,{\times}\, n$ matrix
\begin{equation}\label{M_I}
  \matM_I:=
  \mbox{\standardfootnotesize$\displaystyle
    \addtolength{\arraycolsep}{-2pt}
    \begin{pmatrix}
      \,P_{I}^1\\
      {}&\ddots\\
      {}&{}&P_{I}^{\ n-2r}
      &&&&\makebox[0pt]{\qquad\qquad
        $\smash{\mbox{\Huge${0}$}}$}\\[4pt]
      {}&{}&{}&P_{I}^{\ n-2r+1}\;&P_{I}^{\ n-2r+2}
      &{}&\\
      {}&{}&{}&0&P_{I}^{\ n-2r+1}\\[4pt]
      {}&{}&{}&{}&{}&P_{I}^{\ n-2r+3}\;&P_{I}^{\ n-2r+4}\\
      {}&{}&{}&{}&{}&0&P_{I}^{\ n-2r+3}\\[4pt]
      {}&{}& \makebox[0pt]{\qquad\qquad$\smash{\mbox{\Huge${0}$}}$}
      &{}&{}&{}&{}&\ddots\\
      {}&{}&{}&{}&{}&{}&{}&{}&P_{I}^{\ n-1}&P_{I}^{\ n}\\
      {}&{}&{}&{}&{}&{}&{}&{}&0&P_{I}^{\ n-1}
    \end{pmatrix}\!,$}
\end{equation}
which is the direct sum of a diagonal matrix and $r$ upper-triangular
$\,2\,{\times}\,2$ matrices.  These matrices relate the rows of
$\eigenP$ as
\begin{equation}\label{pi-M}
  \pi_I=\pi_{\Omega}\,\matM_I,\quad I=1,\dots,n.
\end{equation}
They can be \textit{characterized} as the upper-triangular
$2\,{\times}\,2$\,-block-diagonal matrices that satisfy~\eqref{pi-M}.

The next result answers the problem addressed
in~Q\ref{which-canonical}.


\begin{Prop}\label{prop-M} The structure constants are reconstructed
  from the eigenmatrix as
  \begin{equation}\label{N-from-M}
    N_I=\eigenP\, \matM_I\,\eigenP^{-1}.
  \end{equation}
\end{Prop}
\begin{proof} 
  The regular representation $\lambda{:}\;\algF\,{\to}\,\End\vectF$ of
  the algebra~$\algF$, where~$\vectF$ is the underlying vector space,
  is faithful because~$\one\,{\in}\,\algF$; therefore, $\algF$ is
  completely determined by its regular representation.  By definition,
  the matrices $N_I$ represent the elements~$X_I\,{\in}\,\algF$ in the
  basis $X_\bullet$:
  \begin{equation*}
    \lambda(X_I)=N_I.
  \end{equation*}
  On the other hand, using relations~\eqref{ee}\,--\,\eqref{ww}, we
  calculate
  \begin{equation*}
    X_I Y_A=
    \begin{cases}
      P_I^{\ J}Y_J,                  & J\,{=}\,1,2,\dots,n{-}2r,\\
      P_I^{\ J}Y_J+P_I^{\ J+1}Y_{J+1}, & J\,{=}\,n{-}2r{+}2i{+}1,\ i\,{\geq}\,0,\\
      P_I^{\ J-1}Y_J,                & J\,{=}\,n{-}2r{+}2i,\ i\,{\geq}\,1
    \end{cases}
  \end{equation*}
  (no summation over $J$).  This implies that the matrices $\matM_I$
  in~\eqref{M_I} represent the elements~$X_I\,{\in}\,\algF$ in the
  basis $Y_\bullet$, and hence~\eqref{N-from-M} follows.
\end{proof}



\begin{Rem}
  The eigenmatrix $\eigenP$ of the fusion algebra is different from
  the modular transformation matrix $\pureS$ even in the semisimple
  case.  The most essential part of the semisimple Verlinde formula
  consists in the relation between the eigenmatrix~$\eigenP$, which
  maps the canonical basis of the fusion algebra to primitive
  idempotents, and the matrix $\pureS$, which represents
  $S\,{\in}\,\SLiiZ$ on characters,
  \begin{equation}\label{ssP=SK}
    \eigenP=\pureS\,\theK_{\mathrm{diag}},
  \end{equation}
  with $\theK_{\mathrm{diag}}$ in turn expressed through the elements
  $(S_{\Omega}^{\ 1},S_{\Omega}^{\ 2},\dots,S_{\Omega}^{\ n})$ of the
  vacuum row of~$\pureS$,
  \begin{equation}\label{eq:semisimple-K}
    \theK_{\mathrm{diag}}
    :=\diag\bigl(\ffrac{1}{S_{\Omega}^{\ 1}},
    \ffrac{1}{S_{\Omega}^{\ 2}},
    \dots,\ffrac{1}{S_{\Omega}^{\ n}}\bigr).
  \end{equation}
  In the nonsemisimple case, a relation between $\pureS$ and $\eigenP$
  generalizing~\eqref{ssP=SK}\,--\,\eqref{eq:semisimple-K} gives the
  nontrivial part of the corresponding generalized Verlinde formula.
  This is studied in the next subsection.
\end{Rem}

\subsection{From $\pureS$ to $\eigenP$}\label{sec:StoP}
We now construct $\eigenP$ from $\pureS$ via a generalization of the
Verlinde formula to nonsemisimple fusion algebras described
in~\eqref{ee}\,--\,\eqref{ww}.  The first step is to construct the
interpolating matrix~$\theK$ generalizing $\theK_{\mathrm{diag}}$; the
diagonal structure present in the semisimple case is replaced by a
$2\,{\times}\,2$ block-diagonal structure.  We recall that the rows
and columns of $\pureS$ are labeled by representations, and that the
distinguished row
\begin{equation}\label{vac-row}
  \sigma_\Omega=
  (\,S_{\Omega}^{\ 1},\,S_{\Omega}^{\ 2},\,\dots,\,
  S_{\Omega}^{\ 2p-1},\,S_{\Omega}^{\ 2p}\,)\notag
\end{equation}
of~$\pureS(p)$ corresponds to the vacuum representation.  Then $\theK$
is the block-diagonal matrix
\begin{equation}\label{K-blocks}
  \theK
  :=K_0\oplus K_1\oplus\dots\oplus K_{p-1},
  \qquad K_i
  \in\mathrm{Mat}_{2}(\oC)
\end{equation}
with
\begin{equation}\label{build-K}
  K_0
  := \begin{pmatrix}
    \ffrac{1}{S_{\Omega}^{\ 1}}\;&0\\
    0\;&\ffrac{1}{S_{\Omega}^{\ 2}}
  \end{pmatrix}\!,
  \qquad
  K_j
  := \begin{pmatrix}
    \ffrac{1}{S_{\Omega}^{\ 2j+1}-S_{\Omega}^{\ 2j+2}}\;
    &-S_{\Omega}^{\ 2j+2}\\[8pt]
    \ffrac{-1}{S_{\Omega}^{\ 2j+1}-S_{\Omega}^{\ 2j+2}}\;
    &S_{\Omega}^{\ 2j+1}
  \end{pmatrix}
\end{equation}
for $j\,{=}\,1,\dots,p{-}1$.  This matrix relates the distinguished
rows of $\eigenP$ and $\pureS$ as
\begin{equation}\label{sigma-K}
  \pi_\Omega =\sigma_\Omega \,\theK.
\end{equation}
It can be \textit{characterized} as the block-diagonal matrix of
form~\eqref{K-blocks}, with diagonal $K_0$ and with each $K_i$,
$i\,{=}\,1,\dots,p{-}1$, of the form 
$K_i\,{=}\,\smatrix{k_i}{*}{-k_i}{*}$ defined
up to a normalization of the second column, that
satisfies~\eqref{sigma-K}.  In~\eqref{build-K}, we chose the
normalization such that $\det K_j\,{=}\,1$; the freedom in this (nonzero)
normalization factor is related to the freedom in rescaling each
element in the radical, and hence rescaling the corresponding columns
of~$\eigenP$.

For a given $\pureS$, we set (restoring the explicit dependence on the
parameter~$p$ that specifies the model)
\begin{equation}\label{build-P}
  \eigenP(p) :=\pureS(p) \,\theK(p).
\end{equation}
The above prescription for the interpolating matrix $\theK$ and the
resulting expression~\eqref{build-P} for the eigenmatrix $\eigenP$
solve problem~Q\ref{what-K} in the Introduction.


\begin{Rem} Combining formulas~\eqref{N-from-M}
  and~\eqref{build-P}, we can write the generalized Verlinde formula as
\begin{equation}
  N_I=\pureS\,(\theK\,\tildeS_I)\,\pureS^{-1},
\end{equation}
where $\tildeS_I\,{:=}\,\matM_I\theK^{-1}$.  In the semisimple case,
this reduces to the ordinary Verlinde formula written as
  $N_I\,{=}\,\pureS\,(\theK_{\mathrm{diag}}\,
  \tildeS_{\mathrm{diag},I})\,\pureS^{-1}$,
with diagonal matrices $\theK_{\mathrm{diag}}$ given by~\eqref{ssP=SK}
and $(\tildeS_{\mathrm{diag},I})_J^{\ K} \,{=}\, 
\pureS_I^{\ K}\,\delta_J^{\ K}$.
\end{Rem}

In the $(1,p)$ model, we use the $\pureS(p)$ matrix obtained in
Sec.~\ref{sec:char-to-S} and its distinguished row~\eqref{sigma-sharp}
to derive
\begin{multline*}
  K_0
  = p\sqrt{2p}\begin{pmatrix}
    1\;&0\\
    0\;&(-1)^{p+1}
  \end{pmatrix}\!,
  \\
  K_j
  = (-1)^{p+j}\sqrt{\ffrac{p}{2}}
  \begin{pmatrix}
    \ffrac{-1}{\sin\frac{j\pi}{p}} &\ffrac{2}{p^2}
    \bigl(\cos\ffrac{j\pi}{p} - j\sin\ffrac{j\pi}{p}\bigr)\\[10pt]
    \ffrac{1}{\sin\frac{j\pi}{p}}\; &\ffrac{-2}{p^2}
    \bigl(\cos\ffrac{j\pi}{p} + (p{-}j)\sin\ffrac{j\pi}{p}\bigr)
  \end{pmatrix}\!,\quad j\,{=}\,1,\dots,p{-}1.
\end{multline*}
A straightforward calculation then shows that
\begin{equation}\label{P-blocks}
  \eigenP(p)=
  \begin{pmatrix}
    P_{0,0}&P_{0,1}&\dots&P_{0,p-1}\\
    P_{1,0}&P_{1,1}&\dots&P_{1,p-1}\\
    \hdotsfor{4}\\
    P_{p-1,0}&P_{p-1,1}&\dots&P_{p-1,p-1}
  \end{pmatrix}.
\end{equation}
with the $2\,{\times}\,2$ blocks
\begin{alignat}{2}
  P_{0,0}&=
  \begin{pmatrix}
    p\; & (-1)^{p+1}p\\
    p\; & -p
  \end{pmatrix}\!,
  &\quad P_{0,j}&=
  \begin{pmatrix}
    0\;&-\ffrac{2}{p}\sin\ffrac{j\pi}{p}\\[9pt]
    0\;&-(-1)^{j+p}\,\ffrac{2}{p}\sin\ffrac{j\pi}{p}
  \end{pmatrix}\!,
  \\
  P_{s,0}&=
  \begin{pmatrix}
    s\; & (-1)^{s+1}s\\[2pt]
    p{-}s\; & (-1)^{s+1}(p{-}s)
  \end{pmatrix}\!,
\end{alignat}
and
\begin{equation}\label{Psj}
  P_{s,j}=
  (-1)^s
  \begin{pmatrix}
    -\ffrac{\sin\frac{sj\pi}{p}}{\sin\frac{j\pi}{p}}
    &\ffrac{2}{p^2}\Bigl(-s\cos\ffrac{sj\pi}{p}\sin\ffrac{j\pi}{p}
    +\sin\ffrac{sj\pi}{p}\cos\ffrac{j\pi}{p}\Bigr)\\[12pt]
    \ffrac{\sin\frac{sj\pi}{p}}{\sin\frac{j\pi}{p}}
    &\ffrac{2}{p^2}\Bigl(
    -(p{-}s)\cos\ffrac{sj\pi}{p}\sin\ffrac{j\pi}{p}
    -\sin\ffrac{sj\pi}{p}\cos\ffrac{j\pi}{p}\Bigr)
  \end{pmatrix}
\end{equation}
for $s,j\,{=}\,1,\dots,p{-}1$.  

The first column of $\eigenP(p)$ contains the quantum dimensions of
all the irreducible representations in the model. They are given by
\begin{equation*} 
  (\, p,\, p,\, 1,\, p{-}1,\, 2,\, p{-}2,\, \dots,\, p{-}1,\, 1\, ),
\end{equation*}
listed in the order~\eqref{rep-order}, i.e.,
\begin{equation}\label{q-dims'}
  \mathrm{qdim}(\repLambda(s)) = s = \mathrm{qdim}(\repPi(s)),
  \quad s\,{=}\,1,\dots,p.
\end{equation}
Remarkably, all these quantum dimensions are integral.  This points to
an underlying quantum-group structure, such that the quantum
dimensions are the dimensions of certain quantum group modules.  This
quantum-group structure will be considered elsewhere (see more
comments in the Conclusions, however).

As noted above, the normalization of each even column of~$\eigenP$
starting with the fourth can be changed arbitrarily because $w_\alpha$
in the radical cannot be canonically normalized.


\begin{example}
  For $p\,{=}\,2,3,4$, the eigenmatrices found above are evaluated as
  follows:
  \begin{gather*}\addtolength{\arraycolsep}{2pt}
    \eigenP(2)=
    \begin{pmatrix}
      2 & -2 & 0 & 1 \\[1pt]
      2 & -2 & 0 & -1 \\[1pt]
      1 & 1 & 1 & 0 \\[1pt]
      1 & 1 & -1 & 0
    \end{pmatrix}\!,\quad
    \eigenP(3)=\begin{pmatrix}
      3 & 3 & 0 & - \frac{1} {{\sqrt{3}}} & 0
      & \frac{1} {{\sqrt{3}}} \\[1pt]
      3 & -3 & 0 & - \frac{1} {{\sqrt{3}}} & 0 &
      - \frac{1} {{\sqrt{3}}} \\[1pt]
      1 & 1 & 1 & 0 & 1 & 0 \\[1pt]
      2 & 2 & -1 & \frac{1} {2{\sqrt{3}}} & -1 & \frac{-1}
      {2{\sqrt{3}}} \\[1pt]
      2 & -2 & -1 & \frac{1} {2{\sqrt{3}}} & 1 & \frac{1}
      {2{\sqrt{3}}} \\[1pt]
      1 & -1 & 1 & 0 & -1 & 0  \end{pmatrix}\!,\\[5pt]
    \eigenP(4)=
    \mbox{\small$\displaystyle\addtolength{\arraycolsep}{1pt}
      \begin{pmatrix} 4 & -4 & 0 & \frac{1} {2{\sqrt{2}}} & 0
        &
        - \frac{1}{2} & 0 & \frac{1} {2{\sqrt{2}}} \\[3pt]
        4 & -4 & 0 & \frac{-1} {2{\sqrt{2}}} & 0 & - \frac{1}{2} & 0 &
        \frac{-1} {2{\sqrt{2}}} \\[3pt]
        1 & 1 & 1 & 0 & 1 & 0 & 1 & 0 \\[1pt]
        3 & 3 & -1 & \frac{1}{4} & -1 & 0 & -1 & - \frac{1}{4} \\[3pt]
        2 & -2 & -{\sqrt{2}} & \frac{1} {8{\sqrt{2}}} & 0 & \frac{1}{4}
        & { \sqrt{2}} & \frac{1} {8{\sqrt{2}}} \\[3pt]
        2 & -2 & { \sqrt{2}} & \frac{-1} {8{\sqrt{2}}} & 0 & \frac{1}{4}
        & -{ \sqrt{2}} & \frac{-1} {8{\sqrt{2}}} \\[3pt]
        3 & 3 & 1 & - \frac{1}{4} & -1 & 0 & 1 & \frac{1}{4} \\[1pt]
        1 & 1 & -1 & 0 & 1 & 0 & -1 & 0
      \end{pmatrix}\!.$}
  \end{gather*}
\end{example}

\subsection{The fusion algebra $\falgebra_p$}\label{sec:find-fusion}
From $\fancyS_p(\tau)$ in~\eqref{fancyS}, we have arrived at the
eigenmatrix $\eigenP(p)$ in~\eqref{P-blocks}\,--\,\eqref{Psj}.  As we
saw in Sec.~\ref{sec:eigenP}, the fusion is reconstructed from the
eigenmatrix.  We now perform this reconstruction for the $(1,p)$
model.


\begin{Thm}\label{Thm:fusion} For each $p\,{\geq}\,2$, the fusion
  algebra $\falgebra_p$ determined by the eigenmatrix~$\eigenP(p)$ is
  described by the following multiplication table of the $2p$
  canonical basis elements $\repLambda(p)$, $\repPi(p)$,
  $\repLambda(1)$, $\repPi(p{-}1)$, $\repLambda(2)$, $\repPi(p{-}2)$,
  \dots, $\repLambda(p{-}1)$, $\repPi(1)$:
  \begin{alignat*}{2}
    \repLambda(s)\fusion\repLambda(t) &=
    \mathop{\sum_{r=|s-t|+1}}\limits_{\mathrm{step}=2}^{s+t-1}
    \widetilde{\!\repLambda}(r),&\qquad \repLambda(s)\fusion\repPi(t) &=
    \mathop{\sum_{r=|s-t|+1}}\limits_{\mathrm{step}=2}^{s+t-1}
    \widetilde{\!\repPi}(r),\\
    \repPi(s)\fusion\repPi(t) &=
    \mathop{\sum_{r=|s-t|+1}}\limits_{\mathrm{step}=2}^{s+t-1}
    \widetilde{\!\repLambda}(r),
    &s,t&=1,\dots,p,
  \end{alignat*}
  where
  \begin{align*}
    \widetilde{\!\repLambda}(r) &:= \begin{cases}
      \repLambda(r),&1\leq r\leq p,\\
      \repLambda(2p{-}r)+2\repPi(r{-}p),& p{+}1\leq r\leq 2p{-}1,
    \end{cases}\\
    \widetilde{\!\repPi}(r) &:= \begin{cases}
      \repPi(r),&1\leq r\leq p,\\
      \repPi(2p{-}r)+2\repLambda(r{-}p),& p{+}1\leq r\leq 2p{-}1.
    \end{cases}
  \end{align*}
\end{Thm}
\begin{proof} 
  We first evaluate the matrices~$\matM_I$ in accordance
  with~\eqref{M_I}.  For each $s\! = 0,\dots,p{-}1$, the matrix
  $\matM_{2s+1}$ corresponds to the $(2s{+}1)$th row of the
  eigenmatrix~$\eigenP(p)$, and hence to the
  representation~$\repLambda(s)$.  For $s\,{=}\,1,\dots,p{-}1$, we
  have\pagebreak[3]
  \begin{multline*}
    \matM_{2s+1}\equiv \matM(\repLambda(s))=\\*
    \mbox{\standardfootnotesize$\displaystyle
      \addtolength{\arraycolsep}{-1pt}
      \begin{pmatrix}
        \,s\;\\
        &(-1)^{s+1}s\\
        &&\mbox{\Large$\ddots$}\\
        &&&(-1)^{s+1}\ffrac{\sin\frac{sj\pi}{p}}{\sin\frac{j\pi}{p}}\;\;&
        \ffrac{2(-1)^s}{p^2}\Bigl(\sin\frac{sj\pi}{p}\cos\frac{j\pi}{p}
        -s\cos\frac{sj\pi}{p}\sin\frac{j\pi}{p}\Bigr)\\
        &&&0&(-1)^{s+1}\ffrac{\sin\frac{sj\pi}{p}}{\sin\frac{j\pi}{p}}\\
        &&&&&\mbox{\Large$\ddots$\,}
      \end{pmatrix}\!,$}
  \end{multline*}
  where the dots denote the $2\,{\times}\,2$ block of the indicated
  structure written $p{-}1$ times, for $j\,{=}\,1,\dots,p{-}1$.  (In
  particular, $\matM_3\,{=}\,\bs{1}$; the matrices $\matM_1$ and
  $\matM_2$ have a simple form and are not written here for brevity.)
  The matrices $\matM_{2s+2}$, $s\,{=}\,0,\dots,p{-}1$, have a similar
  structure, which can be written most compactly by first noting that
  \begin{gather*}
    \matM_{4}\equiv \matM(\repPi(1))=
    \begin{pmatrix}
      \,1\\
      &(-1)^p\\
      &&\ddots\\
      &&&(-1)^{p+j}&0\\
      &&&0&(-1)^{p+j}\\
      &&&&&\ddots\,
    \end{pmatrix}
  \end{gather*}
  (where the block is again to be written $p{-}1$ times, for
  $j\,{=}\,1,\dots,p{-}1$) and then
  \begin{equation}\label{Pi1-act}
    \matM_{2s+2}\equiv \matM(\repPi(s))
    =\matM(\repPi(1))\,\matM(\repLambda(s)).
  \end{equation}
  
  With the $\matM_I$ matrices thus found, we can reconstruct the
  structure constants from \eqref{N-from-M}.  But it is technically
  easier to find the same structure constants from the algebra
  satisfied by the matrices $\matM_I$,
  \begin{equation*}
    \smash[t]{\matM_I\, \matM_J = \sum_{K=1}^{2p} N_{IJ}^{K}\,\matM_K,}
  \end{equation*}
  which (just by~\eqref{N-from-M}) furnish an equivalent
  representation of the fusion algebra.
  
  {}From~\eqref{Pi1-act}, we conclude that
  $\repPi(1)\,{\fusion}\,\repLambda(s)\,{=}\,\repPi(s)$; it
  immediately follows that
  $\repPi(1)\,{\fusion}\,\repPi(s)\,{=}\,\repLambda(s)$,
  $s\,{=}\,1,\dots,p$.  By associativity, it therefore remains to
  prove only the $\repLambda(s)\,{\fusion}\,\repLambda(t)$ fusion,
  that is, to show the matrix identities (assuming $s\,{\geq}\, t$ for
  definiteness)
  \begin{equation*}
    \matM_{2s+1}\,\matM_{2t+1}=\sum_{a=0}^{t-1}
    \matM(\,\widetilde{\!\repLambda}(s{-}t{+}1{+}2a)),
  \end{equation*}
  where we extend the mapping $\repLambda(s)\,{\mapsto}\,
  \matM(\repLambda(s))$, $\,\repPi(s)\,{\mapsto}\,\matM(\repPi(s))$ by
  linearity, such that
  \begin{equation*}
    \matM(\,\widetilde{\!\repLambda}(r))=\matM(\repLambda(2p{-}r))+
    2\,\matM(\repPi(r{-}p)) =\matM_{2(2p-r)+1}+2\,\matM_{2(r-p)+2}
  \end{equation*}
  for $r\,{\geq}\, p\,{+}\,1$.  But elementary calculations with the
  matrices explicitly given above show that
  \begin{equation*}
    \matM(\,\widetilde{\!\repLambda}(r))=\matM(\repLambda(r))
  \end{equation*}
  (which may be rephrased by saying that
  $\matM(\,\widetilde{\!\repLambda}(r))$ ``continues''
  $\matM(\repLambda(r))$ to $r\,{\geq}\, p{+}1$).  Therefore, the
  statement of the theorem reduces to the matrix identity
  \begin{equation*}
    \matM_{2s+1}\matM_{2t+1}=\sum_{a=0}^{t-1}\matM_{2(s-t+1+2a)+1},
  \end{equation*}
  which can be verified directly.  For the upper-left $2\,{\times}\,2$
  blocks, this is totally straightforward,
  \begin{multline*}
    \begin{pmatrix}
      s\;&0\\
      0\;&(-1)^{s+1}s
    \end{pmatrix}
    \begin{pmatrix}
      t\;&0\\
      0\;&(-1)^{t+1}t
    \end{pmatrix}\\*
    =\sum_{a=0}^{t-1}
    \begin{pmatrix}
      s-t+1+2a&0\\
      0&(-1)^{s+t}(s-t+1+2a)
    \end{pmatrix}\!,
  \end{multline*}
  and for the other blocks the calculation amounts to evaluating sums
  of the form
  \begin{equation*}
    \sum_{a=0}^{t-1}\sin\ffrac{r+2a}{\alpha}=
    \frac{\sin\frac{t}{\alpha}\,\sin\frac{r+t-1}{\alpha}}{
      \sin\frac{1}{\alpha}}
  \end{equation*}
  and their derivatives.
\end{proof}


\begin{Rem}
  We see that $\repPi(1)$ is a simple current of order two, acting
  without fixed points; it underlies the symmetry~\eqref{vee}.  This
  simple current symmetry is analogous to the one present in rational
  CFTs. The permutations of the entries of $\pureS(p)$ correspond to
  the action of the simple current $\Pi(1)$ by the fusion product,
  while the sign factors are exponentiated monodromy charges, which
  are combinations of conformal weights.
  
  We also note that the quantum dimensions \eqref{q-dims'} furnish a
  one-dimensional representation of the fusion algebra.
\end{Rem}


\begin{example}  For $p\,{=}\,2$, the $\falgebra_2$ algebra coincides
  with the fusion obtained in~\cite{GK1}, written in terms of linearly
  independent elements corresponding to the irreducible subquotients,
  as explained above.
  
  For $p\,{=}\,3$ and $4$, we write the fusion algebras explicitly.
  To reduce the number of formulas, we note that for all $p$, \ 
  $\repLambda(1)$ is the unit element and $\repPi(1)$ is an order-$2$
  simple current that acts as
  \begin{equation*}
    \repPi(1) \fusion \repLambda(s) = \repPi(s),\qquad
    \repPi(1) \fusion \repPi(s) = \repLambda(s).
  \end{equation*}
  Further, $\repPi(s)\,{\fusion}\,\repPi(t)\,{=}\,
  \repLambda(s)\fusion\repLambda(t)$ and
  $\repLambda(s)\,{\fusion}\,\repPi(t)\,{=}\,
  \repLambda(t)\fusion\repPi(s)$.  The remaining relations are now
  written explicitly.
  
  For $p\,{=}\,3$, the remaining $\falgebra_3$ relations are given by
  \begin{alignat*}{2}    
    \repLambda(2)\fusion \repLambda(2) &= \repLambda(1)
    + \repLambda(3),&
    \repLambda(2)\fusion \repLambda(3) &=
    2 \repLambda(2) + 2 \repPi(1),\\
    \repLambda(2)\fusion \repPi(2) &= \repPi(1) + \repPi(3),&
    \repLambda(2)\fusion \repPi(3) &=
    2\repPi(2) + 2\repLambda(1),\\
    \repLambda(3)\fusion \repLambda(3) &=
    2 \repLambda(1) + 2 \repPi(2) + \repLambda(3),\\
    \repLambda(3)\fusion \repPi(3) &=
    2 \repLambda(2) + 2 \repPi(1) + \repPi(3),
  \end{alignat*}
  
  For $p\,{=}\,4$, the remaining $\falgebra_4$ relations are
   \begin{alignat*}{2}
     \repLambda(2)\fusion\repLambda(2) &= \repLambda(1)
     + \repLambda(3),&
     \repLambda(2)\fusion\repLambda(3) &= \repLambda(2)
     + \repLambda(4),\\
     \repLambda(2)\fusion\repLambda(4) &= 2\repPi(1)
     + 2\repLambda(3),&
     \repLambda(3)\fusion\repLambda(3) &= \repLambda(1) + 2
     \repLambda(3)
     + 2\repPi(1),\\
     \repLambda(3)\fusion\repLambda(4) &= 2\repLambda(2) + 2 \repPi(2)
     + \repLambda(4),\kern-20pt\\
     \repLambda(4)\fusion\repLambda(4) &= 2\repLambda(1) + 2 \repPi(3)
     + 2\repLambda(3) + 2\repPi(1),\kern-250pt\\
     \repLambda(2)\fusion\repPi(2) &= \repPi(1) + \repPi(3),&
     \repLambda(2)\fusion\repPi(3) &= \repPi(2) + \repPi(4),\\
     \repLambda(2)\fusion\repPi(4) &=
     2\repPi(3) + 2\repLambda(1),&
     \repLambda(3)\fusion\repPi(3) &= \repPi(1) + 2\repPi(3) +
     2\repLambda(1),\\
     \repLambda(3)\fusion\repPi(4) &=
     2\repPi(2) + 2\repLambda(2) + \repPi(4),\kern-20pt\\
     \repLambda(4)\fusion\repPi(4) &= 2\repPi(1) + 2\repLambda(3)
     + 2\repPi(3) + 2\repLambda(1),\kern-250pt
   \end{alignat*}
\end{example}

\section{Conclusions}
To summarize, our proposal for a nonsemisimple generalization of the
Verlinde formula is given by~\eqref{build-P}, with the interpolating
matrix~$\theK$ built in accordance
with~\eqref{vac-row}\,--\,\eqref{build-K} from $\pureS$ constructed
in~\eqref{pureS}. From the matrix~$\eigenP$ that is provided by the
generalized Verlinde formula~\eqref{build-P}, the structure constants
of the fusion algebra are reconstructed via~\eqref{M_I}
and~\eqref{N-from-M}.  In $(1,p)$ models, this leads to the fusion in
Theorem~\ref{Thm:fusion}.

The rest of this concluding section is more a todo list than the
conclusions to what has been done.  First, we have used a
generalization of the Verlinde formula to derive the fusion in $(1,p)$
models, see Theorem~\ref{Thm:fusion}, but we have not presented a
systematic ``first-principle'' proof of the proposed recipe.  The
relevant first principles are the properly formulated axioms of chiral
conformal field theory.  The situation is thus reminiscent of the one
with the ordinary (semisimple) Verlinde formula, whose proof could be
attacked only after those axioms had been formulated~\cite{MS} (see
also~\cite{FbZ,BK}) for rational conformal field theory.  In the
semisimple case, the structure constants are expressed through the
defining data of the representation category, which is a modular
tensor category, and thus through the matrices of the basic $B$ and
$F$ operations of~\cite{MS} as
\begin{equation*}
  \sum_j
  S_{ij}\,
  \frac{\left(
      B\!\left[\begin{smallmatrix}
          j\dual&k\\
          j&k
        \end{smallmatrix}
      \right]
      B\!\left[\begin{smallmatrix}
          k&j\dual\\
          j\dual&k\dual
        \end{smallmatrix}
      \right]
    \right)_{\!00}}{F_k}
  \,S_{jl}=N_{ikl},
\end{equation*}
where
\begin{equation*}
  F_k=F^{}_{00}\!\left[\begin{smallmatrix}
      k\dual&k\\
      k&k
    \end{smallmatrix}
  \right].
\end{equation*}
These formulas are to be related to the above construction of the
fusion algebra constants expressed as
\begin{equation*}
  N_K=\pureS\, \mathscr{O}_K\,\pureS,
\end{equation*}
with the matrices $\mathscr{O}_{I}\,{=}\,\theK \matM_I\theK^{-1}$
(already given in the Introduction) whose structure readily follows
from Sec.~\ref{sec:buld-fusion}.  The necessary modifications of the
RCFT axioms are then to lead to a block-diagonal structure, with
nontrivial blocks being in one-to-one correspondence with the linkage
classes, with the size of a block given by the number of irreducible
representations in the relevant linkage class.

Another obvious task is to place the structures encountered here into
their proper categorical context. For rational CFT, the representation
category $\catC$ of the chiral algebra\,---\,a rational conformal
vertex algebra\,---\,is a modular tensor category, and can thus in
particular be used to associate a three-dimensional topological field
theory to the chiral CFT. For instance, the state spaces of the
three-dimensional TFT are the spaces of chiral blocks of the CFT, and
the modular $S$ matrix (or, to be precise, the symmetric matrix that
diagonalizes the fusion rules) is, up to normalization, the invariant
of the Hopf link in the three-dimensional TFT. Also, a full
(nonchiral) CFT based on a given chiral CFT corresponds to a certain
Frobenius algebra in the category $\catC$, and the correlation
functions of the full CFT can be determined by combining methods from
three-dimensional TFT and from noncommutative algebra in monoidal
categories~\cite{fuRs4,fuRs8}.  In the nonrational case, $\catC$ is no
longer modular, in particular not semisimple, but in any case it
should still be an additive braided monoidal category.  In addition,
other properties of $\catC$, as well as the relevance of
noncommutative algebra in $\catC$ to the construction of full from
chiral CFT, can be expected to generalize from the rational to the
nonrational case.

It is, however, an open (and complicated) problem to make this
statement more precise.  For instance, it is not known how to
generalize the duality structure.  (We note that the fusion rule
algebra $\falgebra_p$ does not share the duality property familiar
from rational fusion algebras: evaluation at the unit element does not
furnish an involution of the algebra.)  On the other hand, the fact
that we are able to identify a finite-dimensional representation of
the modular group in each of the $(1,p)$ models indicates that the
chiral blocks of these models should nevertheless possess the basic
covariance properties under the relevant mapping class group.  This
suggests, in turn, that they can still be interpreted as the state
spaces of a suitable three-dimensional TFT.  (For one proposal on how
to associate a three-dimensional TFT to a nonrational CFT,
see~\cite{KElu}.  However, the $S$ matrix is generically not
symmetric, which certainly complicates the relation to
three-dimensional TFT.)  Furthermore, we expect that this also applies
to many other nonrational CFTs, at least to those for which $\catC$
has a finite number of (isomorphism classes of) simple objects (and
thus in particular finitely many linkage classes), with all of them
having finite quantum dimensions.

A first step in developing the categorical context could consist in
finding the ``fine'' fusion, where each indecomposable $\algW(p)$
representation corresponds to a linearly independent generator in the
fusion algebra.  This fusion would define the monoidal structure of
the category $\catC$.  It should therefore be important for finding
modular invariants and possible boundary conditions in conformal field
theory.  For example, one can imagine that a boundary condition
involves only an indecomposable representation, but \textit{not} its
subquotients (cf.~\cite{brfl,bred}).  A preliminary analysis shows
that for $p\,{=}\,2,3$, invariants
$\bs{\chi}_p^\dagger\,\matH(p)\bs{\chi}_p$ under the $\SLiiZ$ action
on the characters of irreducible $\algW(p)$ representations are given
by
\begin{align*}
  \matH(2)&=
  \begin{pmatrix}
    \frac{1}{4}(h_1+h_2) & 0 & 0 & 0 \\ 0 & 
     \frac{1}{4}(h_1+h_2) & 0 & 
   0 \\ 0 & 0 & h_1 & h_2 \\ 0 & 0 & h_2 & h_1
  \end{pmatrix}\\
  \intertext{and}
  \matH(3)&=
  \mbox{\small$\displaystyle
  \begin{pmatrix}
    \frac{1}{6}(h_1+2 h_2) & 0 & 0 & 0 & 0 & 0 \\
    0 & \frac{1}{6}(h_1+2 h_2) & 0 & 0 & 0 & 0 \\
    0 & 0 & h_1 & h_2 & 0 & 0 \\
    0 & 0 & h_2 & \half(h_1+h_2) & 0 & 0 \\
    0 & 0 & 0 & 0 & \half(h_1+h_2) & h_2 \\
    0 & 0 & 0 & 0 & h_2 & h_1
  \end{pmatrix}\!,$}
\end{align*}
where in each case, the coefficients $h_{1,2}$ must be chosen such
that the matrix entries are integers, for example,
$h_1\,{=}\,h_2\,{=}\,2$ for $p\,{=}\,3$.  The ``fine'' fusion is
needed precisely here in order to correctly interpret the result.  It
allows distinguishing between inequivalent representations that
possess identical characters and is therefore needed for interpreting
the result for the modular invariant as a proper partition function
not only at the level of characters, but also at the level of
representations (or, rephrased in CFT terms, not just describing the
dimensions of spaces of states of the full CFT, but completely telling
which bulk fields result from combining the two chiral parts of the
theory).
 
We also note that behind the scenes in Theorem~\ref{Thm:fusion} is a
quantum group of dimension~$2p^3$.  Its representation category is
equivalent to the category of $\algW(p)$ representations described in
Sec.~\ref{rep-exts}, and the quantum dimensions~\eqref{q-dims'} are
the dimensions of its representations.  The close relation between
this quantum group and the fusion will be studied elsewhere.

Next, the structure of the indecomposable $\algW(p)$ modules in
Sec.~\ref{rep-exts} should be studied further.  This can be done by
traditional means, but a very useful approach is in the spirit
of~\cite{FFHST} (which provides the required description for
$p\,{=}\,2$).  The idea is to add extra modes to the algebra of $a^+$
and $a^-$ in Sec.~\ref{sec:aa} such that the $\algW(p)$ action in the
indecomposable modules is realized explicitly.  With these extra modes
added, some states that are not singular vectors in the module in
Fig.~\ref{fig:LambdaPi} become singular vectors built on new states,
and the construction of these new states can be rephrased as the
``inversion'' of singular vector operators, similarly to how the
operator of the simplest singular vector $L_{-1}$ was inverted
in~\cite{FFHST} (where both the singular vector operator was the
simplest possible and the $a^\pm$ operators were actually
fermions).

Finally, it is highly desirable, but apparently quite complicated, to
extend the analysis in this paper to logarithmic extensions of the
$(p',p)$ models with coprime $p',p\,{\geq}\,2$.  The extended Kac
table of size $p'\,{\times}\,p$ is then selected as the kernel of the
appropriate screening operator.  Already the $(2,3)$ model (which is
trivial in its nonlogarithmic version) is of interest because of its
possible relation to percolation.  However, it is not obvious how to
describe the kernel of the screening in reasonably explicit terms; in
particular, we do not know good analogues of the
operators~$a^+$~and~$a^-$.

\vspace*{4mm}

\subsubsection*{Acknowledgments}
We are grateful to B.~Feigin for useful discussions.  This work was
supported in part by a grant from The Royal Swedish Academy of
Sciences.  AMS and IYuT were also supported in part by the RFBR
(grants 01-01-00906 and LSS-1578.2003.2), INTAS (grant 00-00262), and
the Foundation for Support of Russian Science.  JF and SH are
supported in part by VR under contracts no.\ F\,5102\,--\,2000-5368
and 621\,--\,2002-4226.

\end{document}